\documentclass[a4paper,twocolumn,11pt,unpublished,accepted=2021-06-01]{quantumarticle}
\pdfoutput=1
\usepackage[utf8]{inputenc}
\usepackage[english]{babel}
\usepackage[T1]{fontenc}
\usepackage{amsmath}
\usepackage[sort&compress,numbers]{natbib}
\usepackage{hyperref}

\usepackage{tikz}
\usepackage{lipsum}

\usepackage{bbold}

\newcommand{\ketbra}[2]{\vert #1 \rangle \langle #2 \vert}
\newcommand{\tr}{\text{\normalfont Tr}}

\usepackage{soul} 

\begin{document}

 \title{Certifying dimension of quantum systems by sequential projective measurements}

\author{Adel Sohbi}
\email{sohbi@kias.re.kr}
\affiliation{School of Computational Sciences, Korea Institute for Advanced Study, Seoul 02455, Korea}
\orcid{0000-0002-1275-9722}

\author{Damian Markham}
\affiliation{LIP6, CNRS, Universit\'e Pierre et Marie Curie, Sorbonne Universit\'es, 75005 Paris, France}
\affiliation{JFLI, CNRS, National Institute of Informatics, University of Tokyo, Tokyo, Japan}

\author{Jaewan Kim}
\affiliation{School of Computational Sciences, Korea Institute for Advanced Study, Seoul 02455, Korea}

\author{Marco T\'ulio Quintino}
\email{mtcq.mm@gmail.com}
\affiliation{Vienna Center for Quantum Science and Technology (VCQ), Faculty of Physics,
University of Vienna, Boltzmanngasse 5, 1090, Vienna, Austria}
\affiliation{Institute for Quantum Optics and Quantum Information (IQOQI), Austrian
Academy of Sciences, Boltzmanngasse 3, A-1090 Vienna, Austria}

\maketitle

\begin{abstract}
	This work analyzes correlations arising from quantum systems subject to sequential projective measurements to certify that the system in question has a quantum dimension greater than some $d$. We refine previous known methods and show that dimension greater than two can be certified in scenarios which are considerably simpler than the ones presented before and, for the first time in this sequential projective scenario, we certify quantum systems with dimension strictly greater than three. We also perform a systematic numerical analysis in terms of robustness and conclude that performing random projective measurements on random pure qutrit states allows a robust certification of quantum dimensions with very high probability.
\end{abstract}

\section{Introduction}

With the recent development of quantum technologies and the different promising applications, it is essential to guarantee the good functioning of the used apparatus through certification or benchmarking methods \cite{Eisert2020, kliesch2020theory}. Such methods can rely on fundamental properties of quantum physics to assert properties of quantum systems such as self-testing \cite{Supic2020selftestingof, PhysRevLett.122.250403, PhysRevA.98.062307, PhysRevA.100.022117, Saha_2020}, randomness certification \cite{PhysRevLett.118.060503, Acin2016, PhysRevLett.114.150501} and dimension witnesses \cite{Brunner_2008,PhysRevLett.105.230501, PhysRevLett.110.150501, GBCKL:pra14, PhysRevA.94.032114, Cai_2016, Li2018, Saha_2019}.

The notion of dimension can be defined in abstract ways such as ``the maximal number of perfectly distinguishable states" or, as we do here, simply as the dimension of the quantum system. It has been proved that the usage of qudits instead of qubits is beneficial in a large range of applications in quantum information such as fault-tolerant quantum computation \cite{PhysRevX.2.041021,PhysRevLett.113.230501,PhysRevLett.123.070507}, quantum algorithms \cite{nphys1150, PhysRevA.94.042307, PhysRevA.96.012306}, quantum error correction \cite{DP:pra13,MSBASJG:prx16,GKWYZ:ieee18}, quantum simulation \cite{Neeley722}, universal optics-based quantum computation \cite{NCS:prl18}, quantum communication \cite{CDBO:onlinelibrary19, PhysRevLett.123.070505} and secret sharing \cite{PhysRevA.88.042332}.

In order to certify dimension of single quantum systems, one can use outcomes statistics from a realized experiment in a specific scenario relying on sequential measurements \cite{GBCKL:pra14, Hoffmann_2018, Spee_2020, mao2020structure} or contextuality \cite{PhysRevA.94.032114, sohbi2020experimental, ray2020graphtheoretic}. However, the dimension witnesses in \cite{GBCKL:pra14, PhysRevA.94.032114} are particular cases that are difficult to extend to general cases due to their complexity and the method in \cite{ray2020graphtheoretic} uses rank constrained optimization over positive semidefinite matrices that uses heuristic method which does not guarantee the optimality of the objective function. Another direction is to use the so-called Navascu\'es-Pironio-Ac\'in (NPA) hierarchy \cite{NPA:njp08, PNA:siam10} which is a numerical method initially developed to construct a hierarchy of outer approximations which possibly converge to the set of quantum correlations on a Bell nonlocality scenario. More recently, the NPA hierarchy has been adapted to characterize correlations arising quantum systems with a fixed finite dimensional with the so-called Navascues-Vert\'esi (NV) hierarchy \cite{NV:prl15,NFAV:pra15} (see also \cite{jee2020characterising}) providing computational tools to certify quantum dimensions in a Bell nonlocality and prepare-and-measure scenario \cite{NFAV:pra15}.

In this work we present a systematic computational method to certify dimension of quantum systems subjected to sequential  projective measurements which is based on the finite dimensional NPA hierarchy \cite{NV:prl15,NFAV:pra15}. We apply our methods to obtain the required robustness on experiments which may be subjected to noise, identify good scenarios for certifying a particular desired dimension, and show that certifying quantum dimensions by preforming random measurements on random states can be done with considerably high probability. Finally identify the different advantages to use more experimentally challenging scenarios in order to have a more accurate and robust dimension certification.

This paper is structured as follows.
In Sec.~\ref{sec:bfdsm}, we present preliminary notions on the numerical methods used to certify dimension of quantum systems by sequential projective measurements.
In Sec.~\ref{sec:mnp}, we propose a method to classify scenarios based on the NPA hierarchy and make a proposal to determine what scenario can be used based on this classification. We introduce the notion of robustness that is used to provide a certificate for the dimension.
In Sec.~\ref{sec:mr}, we study different scenarios and compare them to each other in the perspective of dimension certification. In particular, we identify a trade off between the experimental challenges to perform a scenario and a more robust dimension certification. Finally, we show how dimension witness can be computed from our method and provide numerical examples and show that randomly chosen measurements can be used to certify dimension in specific scenario.

\section{Bounding Finite Dimension in a Sequential Projective Measurements Scenario}\label{sec:bfdsm}
\subsection{Sequential Projective Measurements Scenario}\label{sec:sms}

Consider a quantum state $\rho$ in a finite Hilbert space $\mathcal{H}_d \cong \mathbb{C}_d$ which will be subjected to sequential projective measurements described by the projectors $\Pi_{r_i|s_j}$.
Each measurement has an input (or \textit{setting}) $s\in S$ and an output (or \textit{result}) $r\in R$ and the input and output of the $i$-th measurement will be denoted by $s_i$ and $r_i$.

We call an \textit{event} a representation of an output for a specific input. For instance for a single measurement, an event is written as $r_i \vert s_i$ which represents obtaining the output $r_i$ for the input $s_i$ of the $i$-th measurement. Each event is associated to a projector and for the event $r_i \vert s_i$ the associated projector is $\Pi_{r_i\vert s_i}$. Such projectors verify the following conditions, the \textit{completeness} condition: $\sum_{r_i} \Pi_{r_i\vert s_i} = \mathbb{1}$, where $\mathbb{1}$ is the identity matrix and the \textit{orthogonality} condition $\Pi_{r_i\vert s_i}\Pi_{r_i'\vert s_i} = 0$ when $r_i \neq r_i'$.

For events with multiple measurements the ordering is important as it forms a sequence, we label from left to right in order of the sequence, for instance, the event $r_i,r_j\vert s_i, s_j$ which represents obtaining the output $r_i$ for the input $s_i$ for the $i$-th measurement, then obtaining the output $r_j$ for the input $s_j$ for the $j$-th measurement.

\begin{figure}[ht]
  \centering
  \includegraphics[width=8cm]{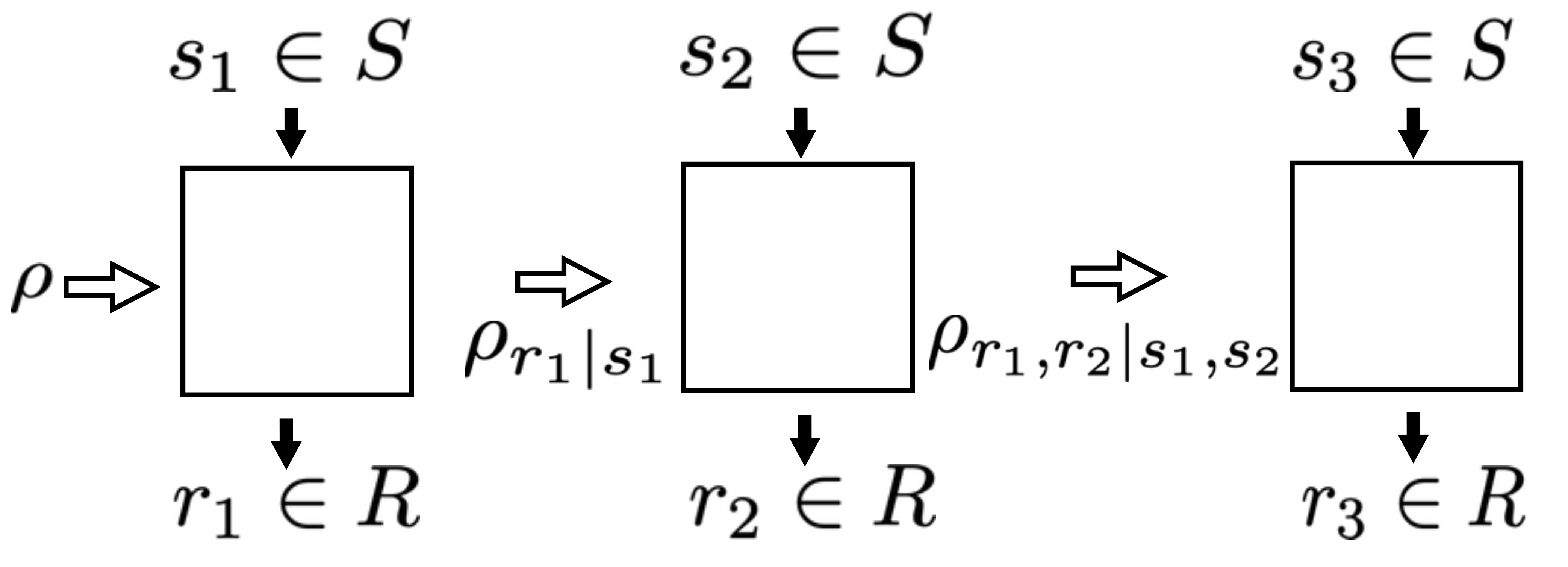}
  \caption{Sequential measurements scenario. Each measurement has an input (or \textit{setting}) $s\in S$ and an output (or \textit{result}) $r\in R$ and the input and output of the $i$-th measurement will be denoted by $s_i$ and $r_i$. }
  \label{fig:seq_meas_sce}
\end{figure}

After the first measurement on the state $\rho$, the post-measured state is denoted by
\begin{equation}
	\rho_{r_1\vert s_1}:= \Pi_{r_1|s_1} \rho \Pi_{r_1|s_1}^\dagger/\text{Tr}[\Pi_{r_1|s_1} \rho \Pi_{r_1|s_1}^\dagger]
\end{equation}
 where $\text{Tr}[.]$ denotes the trace and $\dagger$ is the Hermitian conjugate. After a second measurement, the state is denoted as
 \small
\begin{equation}
\rho_{r_1, r_2\vert s_1, s_2}:=\Pi_{r_2|s_2} \rho_{r_1|s_1} \Pi_{r_2|s_2}^\dagger/\text{Tr}[\Pi_{r_2|s_2} \rho_{r_1|s_1} \Pi_{r_2|s_2}^\dagger].
\end{equation}
\normalsize
	Fig.~\ref{fig:seq_meas_sce} illustrates the case of three sequential measurements on the state $\rho$.

This corresponds to the L\"uders' rule for state update \cite{L_ders_2006}. Other dimension witnesses \cite{Budroni_2014, Schild_2015} have been built based on the von Neumann's rule for state update \cite{Guth1933}.

For simplicity we denote the sequence of outputs $\mathbf{r} = (r_1,\dots,r_l)$ and the sequence of settings $\mathbf{s} = (s_1,\dots,s_l)$. The measurement operator (composition of projectors is not a projector) associated to the event $\mathbf{r}\vert\mathbf{s}$ is $\Pi_{\mathbf{r}\vert\mathbf{s}} = \Pi_{r_l\vert s_l}\dots\Pi_{r_1\vert s_1}$.

The probability to have the event $\mathbf{r}\vert\mathbf{s}$ is given by the Born's rule:

\begin{align}
  P(\mathbf{r}\vert\mathbf{s}) = \text{Tr}[\Pi_{\mathbf{r}\vert\mathbf{s}}\rho \Pi_{\mathbf{r}\vert\mathbf{s}}^\dagger],
\end{align}

In what follows, we consider the case where the set of measurements to choose from is the same for each measurement through the sequence. Moreover, we consider the case where all measurements have the same number of outcomes. Under such assumptions, a scenario is characterized by three following parameters:

\begin{itemize}
  \item The number of measurements: $m$.
  \item The maximum length of sequence of measurements: $l$.
  \item The number of outcomes for each measurements: $o$.
\end{itemize}

We call the MLO (Measurement, Length, Outcomes) representation of a scenario the triplet: $m$-$l$-$o$. As an example the Leggett-Garg scenario \cite{LG:prl85} correspond to the 3-2-2 scenario.

We call a \textit{behavior},
$\mathbf{P}:=\{ P(\mathbf{r}\vert\mathbf{s})  \}_{\mathbf{rs}}$
 , the probability distribution over all the possible events in given sequential measurement scenario. In what follows, we address the issue of certifying dimension by their behaviors. In other words, we are interested to know whether for a behaviour represented by an outcomes statistic $P(\mathbf{r}\vert\mathbf{s})$, there exist a $d$-dimensional state $\rho$ acting in $\mathcal{H}_d$ and a set of projective measurements $\{\Pi_{{r_i}\vert s_i}\}_{i}$ such that $P(\mathbf{r}\vert\mathbf{s}) = \text{Tr}(\Pi_{\mathbf{r}\vert\mathbf{s}} \rho \Pi_{\mathbf{r}\vert\mathbf{s}}^\dagger)$. For a given projective sequential measurement scenario, the set of all quantum behaviors arising from $d$-dimensional systems is described by $\mathcal{Q}_d$.
 
The main questions addressed in this paper can then be described as:
\begin{itemize}
  \item \emph{Feasibility problem}: For a given behavior $P(\mathbf{r}\vert\mathbf{s})$, verify if  $P(\mathbf{r}\vert\mathbf{s})\in\mathcal{Q}_d$, that is, the existence of a $d$-dimensional quantum realization. This problem will be further described in Sec.~\ref{sec:prop} where we also discuss a robust version of it.
  \item \emph{Optimization problem}: Given a set of real coefficient, $\gamma_{\mathbf{r}\vert\mathbf{s}}$, what is the maximum value of $\sum_{\mathbf{r},\mathbf{s}} \gamma_{\mathbf{r}\vert\mathbf{s}} P(\mathbf{r}\vert\mathbf{s})$ under the constraint  $P(\mathbf{r}\vert\mathbf{s})\in\mathcal{Q}_d$, that is, the behavior $P(\mathbf{r}\vert\mathbf{s})$ admits a a $d$-dimensional quantum realization.
\end{itemize}
As we discuss in the next sections, there are methods which allow us to tackle these problems by means of semidefinite programming (SDP) \cite{BMKG:prl13,M:prl90,P:pra90}.

\subsection{Unrestricted Dimensional case}\label{sec:inf}

We first present the study of behaviors of quantum systems in a sequential projective measurements scenario without dimensional constraints, that is, the quantum systems are not restricted to any fixed dimension $d$ \cite{BMKG:prl13}. To this end we introduce the so-called moment matrix representation. A moment matrix $M$ is a symmetric square matrix whose entries, in this case, are all the expectation values of the product of pairs of $\Pi_{\mathbf{r}\vert\mathbf{s}}$. We adopt the specific notation introduced in \cite{BMKG:prl13}, where in the set of projectors, for each setting $\mathbf{s}$, one of the results $\mathbf{r}$ is left out and we add the identity matrix to the list of projectors. From \cite{BMKG:prl13}, the matrix elements of the moment matrix $M$ are:

\begin{align}\label{eq:moment_matrix}
  M_{\mathbf{r}\vert\mathbf{s},\mathbf{r}'\vert\mathbf{s}'} :=  \text{Tr}[\Pi_{\mathbf{r}\vert\mathbf{s}}^\dagger\Pi_{\mathbf{r}'\vert\mathbf{s}'}\rho],
\end{align}
where we use the notation $\Pi_{\mathbf{0}\vert\mathbf{0}} = \mathbb{1}$ to include the identity matrix. Note that $M_{\mathbf{r}\vert\mathbf{s},\mathbf{r}\vert\mathbf{s}} = P(\mathbf{r}\vert\mathbf{s})$.

As pointed in \cite{BMKG:prl13}, it follows from the Born's rule that the moment matrix from a behaviour with a quantum realization is positive semidefinite ($M\geq0$) and that 
\begin{equation} \label{eq:moment_condition}
	M_{\mathbf{r}\vert\mathbf{s},\mathbf{r}'\vert\mathbf{s}'} = M_{\mathbf{r}''\vert\mathbf{s}'',\mathbf{r}'''\vert\mathbf{s}'''}
\end{equation}
 when $\Pi_{\mathbf{r}\vert\mathbf{s}}^\dagger\Pi_{\mathbf{r}'\vert\mathbf{s}'} = \Pi_{\mathbf{r}''\vert\mathbf{s}''}^\dagger\Pi_{\mathbf{r}'''\vert\mathbf{s}'''}$. Also, since we are considering projective measurements we have that $\Pi_{r_i|s_i}\Pi_{r'_{i}|s_j}=\delta_{r_i,r'_{i}}\Pi_{r_i|s_i}$, this will reflect into additional linear constraints in the moment matrix $M$. For instance, the orthogonality constraint ensures entries of the moment matrix to be zero, and the idempotency constraint ($\Pi_{r_i|s_i}\Pi_{r_i|s_i}=\Pi_{r_i|s_i}$) ensures that some entries of moment matrix are identical.
Moreover, when no restrictions in the dimension are imposed, the moment matrices constraints can be used to completely characterize the set of sequential quantum behaviours. More precisely, Ref.\,\cite{BMKG:prl13} exploited the methods of \cite{NPA:njp08,PNA:siam10} to prove that every positive semidefinite operator $M$ respecting the linear conditions imposed by projective measurements ($\Pi_{r_i|s_i}\Pi_{r'_{i}|s_i}=\delta_{r_i,r'_{i}}\mathbb{1}$) and  $M_{\mathbf{0}\vert\mathbf{0},\mathbf{0}\vert\mathbf{0}} = 1 $ has a quantum realization.

When the dimension is unrestricted for any given set of real numbers $\{\gamma_{\mathbf{r}\vert\mathbf{s}}\}$, the \emph{optimization problem} can be solved by the following SDP:

\begin{subequations}
  \label{eq:SDPdiminf}
\begin{alignat}{2}
p^* := &\underset{M}{\text{maximize}}        &\qquad& \sum_{\mathbf{r},\mathbf{s}} \gamma_{\mathbf{r}\vert\mathbf{s}} P(\mathbf{r}\vert\mathbf{s})\label{eq:Mmax}\\
&\text{subject to} &      &  P(\mathbf{r}\vert\mathbf{s}) = M_{\mathbf{r}\vert\mathbf{s},\mathbf{r}\vert\mathbf{s}} \\
&     &      & M_{\mathbf{0}\vert\mathbf{0},\mathbf{0}\vert\mathbf{0}} = 1 ,\label{eq:Mid}\\
&     &      & M \succeq 0,\label{eq:Mpos}\\
&     &      & M_{\mathbf{r}\vert\mathbf{s},\mathbf{r}'\vert\mathbf{s}'} = M_{\mathbf{r}''\vert\mathbf{s}'',\mathbf{r}'''\vert\mathbf{s}'''}  \text{,} \notag\\
&     &      & \text{if } \Pi_{\mathbf{r}\vert\mathbf{s}}^\dagger\Pi_{\mathbf{r}'\vert\mathbf{s}'} = \Pi_{\mathbf{r}''\vert\mathbf{s}''}^\dagger\Pi_{\mathbf{r}'''\vert\mathbf{s}'''} \label{eq:Mlinconst} \\
&     &      & M_{\mathbf{r}\vert\mathbf{s},\mathbf{r}'\vert\mathbf{s}'} = 0  \text{,} \notag\\
&     &      & \text{if } \Pi_{\mathbf{r}\vert\mathbf{s}}^\dagger\Pi_{\mathbf{r}'\vert\mathbf{s}'} = 0. \label{eq:Morthconst}
\end{alignat}
\end{subequations}

This approach can be used to verify the possible quantumness of a behavior regardless of the dimension of its quantum realization.

\subsection{Finite Dimensional case}\label{sec:fin}

The problem of characterising quantum behaviours of some fixed dimension can be tackled by a hierarchy of semidefinite programming relaxations (NV hierarchy) \cite{NV:prl15,NFAV:pra15}, which is inspired by the NPA hierarchy \cite{NPA:njp08, PNA:siam10}. Now, instead of a single semidefinite program to decide weather a given behaviour admits a quantum realization, we define a hierarchy of outer approximations which converges to $\mathcal{Q}_d$. 

The moment matrix associated to  a $d$-dimensional quantum state $\rho_d$ and to length $l$ sequence of $d$-dimensional projective measurements, $\Pi_{\mathbf{r}\vert\mathbf{s}}$ is defined as
\begin{align}\label{eq:moment_matrix_finite}
  M_{\mathbf{r}\vert\mathbf{s},\mathbf{r}'\vert\mathbf{s}'} :=  \text{Tr}[\Pi_{\mathbf{r}\vert\mathbf{s}}^\dagger\Pi_{\mathbf{r}'\vert\mathbf{s}'}\rho_d]
\end{align} 
 and we denote by $\mathcal{M}_d$ the set of all $d$-dimensional length $l$ moment matrices.
In this way, we can write the optimization problem 
\begin{subequations}
  \label{eq:SDPdimfin}
\begin{alignat}{2}
p_{d}^* := &\underset{M}{\text{maximize}}        &\qquad& \sum_{\mathbf{r},\mathbf{s}} \gamma_{\mathbf{r}\vert\mathbf{s}}  P(\mathbf{r}\vert\mathbf{s})\label{eq:Mmaxk}\\
&\text{subject to} &      & P(\mathbf{r}\vert\mathbf{s}) = M_{\mathbf{r}\vert\mathbf{s},\mathbf{r}\vert\mathbf{s}} ,\\
&            &      & M \in \mathcal{M}_d \label{eq:Mdk}.
\end{alignat}
\end{subequations}		
	 Note however that characterizing the set $\mathcal{M}_d$ is not a simple task, hence this formulation does not simplify the problem. One way to obtain a non-trivial relaxation consists of noticing that if $\{M_i\}_i$ be a basis for the linear span of $\mathcal{M}_d$, all elements $M\in\mathcal{M}_d$ have to satisfy
\begin{align}
	&M\geq0 \label{eq:1} \\
	&M=\sum_i \alpha_i M_i,  \quad (\alpha_i\in\mathbb{R}) \label{eq:2} \\
	&M_{\mathbf{0}\vert\mathbf{0},\mathbf{0}\vert\mathbf{0}} = 1 	\label{eq:3}
\end{align}
	Hence, the constraints given by Eq.~\eqref{eq:1}, Eq.~\eqref{eq:2}, and Eq.~\eqref{eq:3} provide an outer approximation of $\mathcal{M}_d$. Also, such relaxation can be made tighter by defining the $k$-th level of the set of moment matrix $\mathcal{M}_d^k$ as the set of moments corresponding to $l+k-1$ measurements. In this way, we have $\mathcal{M}_d=\mathcal{M}_d^{k=1}$ and a tighter approximation can be made by increasing $k$.

\begin{figure}[ht]
  \centering
  \includegraphics[width=6cm]{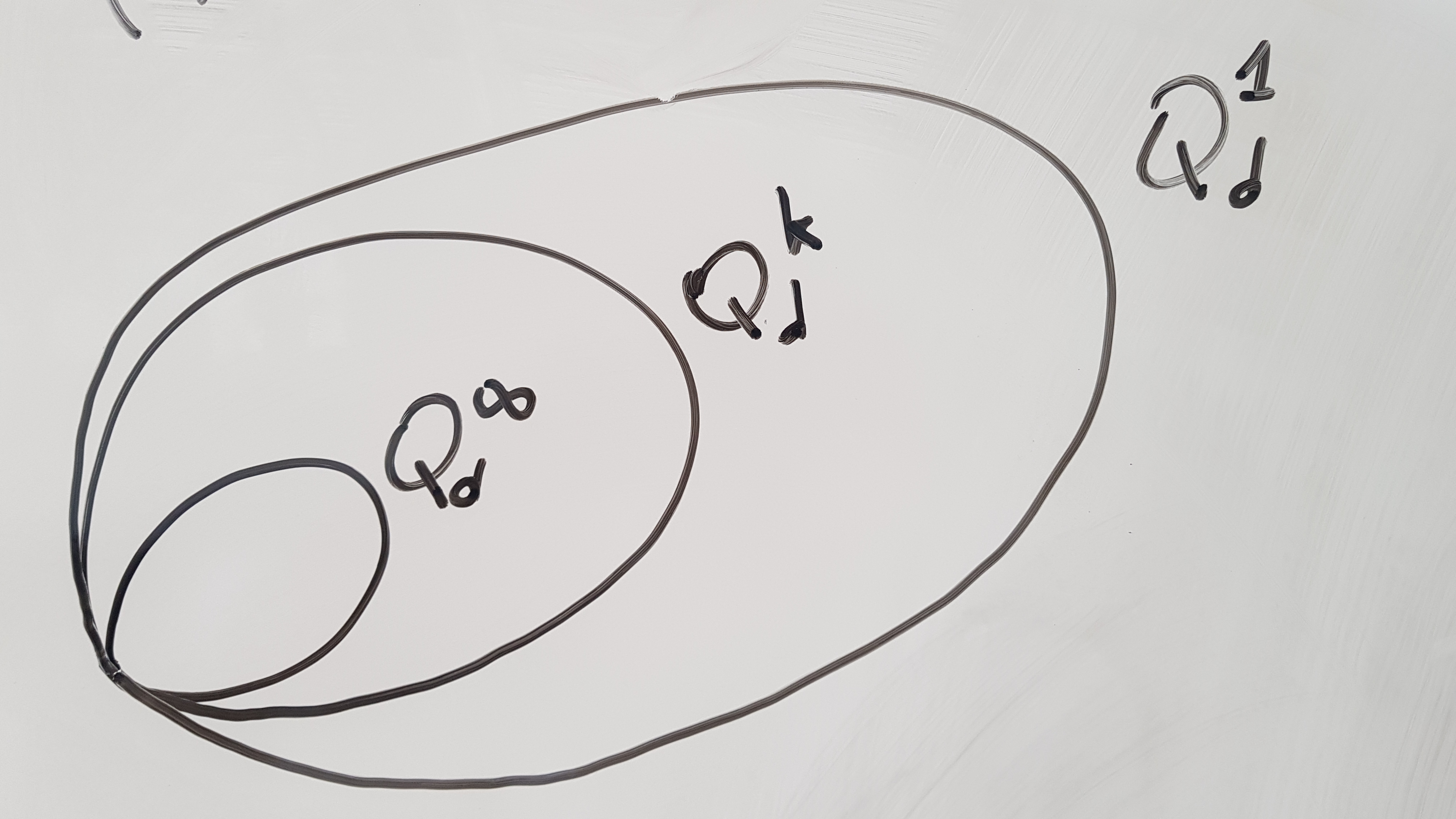}
  \caption{Schematic representation of the sets of behaviors in the different level of the hierarchy such that $\mathcal{Q}_{d}^\infty \subseteq \mathcal{Q}_{d}^k \subseteq \mathcal{Q}_{d}^1$. A behavior $\mathbf{P}$ is given, with $\mathbf{P} \notin \mathcal{Q}_{d}^1$. We have represented the visibility $\eta$ computed with the generalized robustness (see Sec.~\ref{sec:prop}).}
  \label{fig:behaviors_hierachy}
\end{figure}

	If the set $\{M_i\}_i$ is a basis for the linear span of $\mathcal{M}_d^k$, a non-trivial upper bound for the problem presented in Eq.\eqref{eq:SDPdimfin} can be written as \cite{NV:prl15,NFAV:pra15} 
\begin{subequations}
  \label{eq:SDPdimfin2}
\begin{alignat}{2}
p_{d,k}^* := &\underset{M}{\text{maximize}}        & \qquad& \hspace*{-7mm} \sum_{\mathbf{r},\mathbf{s}} \gamma_{\mathbf{r}\vert\mathbf{s}} P(\mathbf{r}\vert\mathbf{s}) = M_{\mathbf{r}\vert\mathbf{s},\mathbf{r}\vert\mathbf{s}} \\
&\hspace*{-10mm}\text{subject to} &      & \hspace*{-10mm} P(\mathbf{r}\vert\mathbf{s}) = M_{\mathbf{r}\vert\mathbf{s},\mathbf{r}\vert\mathbf{s}} \label{eq:c1} \\
&            &      &\hspace*{-10mm} M \succeq 0, \label{eq:c2} \\
&            &      & \hspace*{-10mm}M=\sum_i \alpha_i M_i,  \; (\alpha_i\in\mathbb{R}) \label{eq:c3} \\
&            &      &\hspace*{-10mm} M_{\mathbf{0}\vert\mathbf{0},\mathbf{0}\vert\mathbf{0}} = 1 \label{eq:c4}.
\end{alignat}
\end{subequations}
	In this formulation, each level $k$ of the hierarchy provides a bound for the problem defined in Eq.~\ref{eq:SDPdimfin} such that $p_{d,k+1}^* \geq p_{d,k}^* \geq  p_{d}^*$. Moreover, this hierarchy converges to the actual maximum \cite{NV:prl15,NFAV:pra15}, that is,
	\begin{equation}
	\lim_{k\to\infty} p_{d,k}^*= p_{d}^*.
\end{equation}	 
	We then define $\mathcal{Q}_{d}^k$ as the set of behaviors which respect the constraints in Eq.~\eqref{eq:c1}, Eq.~\eqref{eq:c2}, Eq.~\eqref{eq:c3}, and Eq.~\eqref{eq:c4} and for any level $k$ we are ensured to have an outer approximation $\mathcal{Q}_{d} \subseteq \mathcal{Q}_{d}^k$ (as illustrated in Fig.~\ref{fig:behaviors_hierachy}).

In \cite{NFAV:pra15}, it was shown that for some fixed level $k$ of the hierarchy the set $\mathcal{Q}_{d}^k$ may be strictly larger than $\mathcal{Q}_{d}$. Here we illustrate this fact by presenting a different example. In the guess-your- neighbor’s-input inequality \cite{ABBAGP:prl10}, in particular in its sequential measurement scenario \cite{BMKG:prl13}, it is a sequential scenario with a MLO of 2-3-2 and the inequality is:
\begin{align} \label{eq:gyni}
  P(000|000) + P(110|011) + P(011|101) \notag \\
  +P(101|110) \leq 1.
\end{align}
Reference \cite{BMKG:prl13} evaluated the SDP in Eq.~\ref{eq:SDPdiminf} to find the maximum value $p_2^* \approx 1.0225$ when the dimension is unrestricted. Here we have used the methods discussed in \cite{NFAV:pra15} to obtain  $p_{2,1}^* \approx 1.1588$ for the maximum value for the first level of the hierarchy for a qubit system. This provides a direct proof that there exist moment matrices in the first level of the hierarchy that admit no qubit realization in the scenario 2-3-2 at the first level of the hierarchy for a qubit. This also shows that in the scenario 2-3-2, the first level of the hierarchy for a qubit, some behaviors do not admit quantum realization. Hence, the first level of the hierarchy is not sufficient to fully characterize the set $\mathcal{Q}_d$ of sequential quantum correlations. This example is further discussed in the Appendix~\ref{app:gyni} and the code is provided in \cite{github}.

\section{Methods and Proposal}\label{sec:mnp}

	In order to solve the optimization problem described in Eq.~\ref{eq:SDPdimfin}, one needs first to obtain a basis for the linear span of $\mathcal{M}_d^k$. In particular, if the set $\{M_i\}_i$ forms a basis for the linear span of $\mathcal{M}_d^k$, the problem presented in Eq.\eqref{eq:SDPdimfin} can be tackled by means of semidefinite programming.

  In \cite{NV:prl15,NFAV:pra15}, the authors proposed a randomized method to find a basis for the linear span of $\mathcal{M}_d^k$. The idea is simply generate several random $d$-dimension moment matrices by choosing random measurements and states. But, as stated by the authors, this method has the following problem: since we do not know the number of elements in the basis, we cannot be sure that we have built a complete basis. Here we solve this problem by providing a simple but convincing empirical method that ensures that the random procedure obtained a basis of $\mathcal{M}^k_d$ as wanted.

\subsection{Building a Basis of Moment Matrices}\label{sec:bbmm}

By using the randomized method first proposed in  Ref.\,\cite{NV:prl15,NFAV:pra15} and fully described in Append.~\ref{app:grmm}, it is possible to construct multiple moment matrices from which a basis can be constructed. As pointed in Ref.,\,\cite{NV:prl15,NFAV:pra15}, this method does not ensure that we have completed the processes of finding a basis. We now provide a simple but strong evidence that this is the case. This can be done either by finding the highest number of linearly independent (LI) moment matrices or by using the Gram-Schmidt process on a set of moment matrices until the zero matrix is left through the process. We use the second method while keeping the number of linearly independent moment matrices to characterize a scenario for a specific dimension and level in the hierarchy.

\begin{figure}[ht]
  \centering
  \includegraphics[width=8cm]{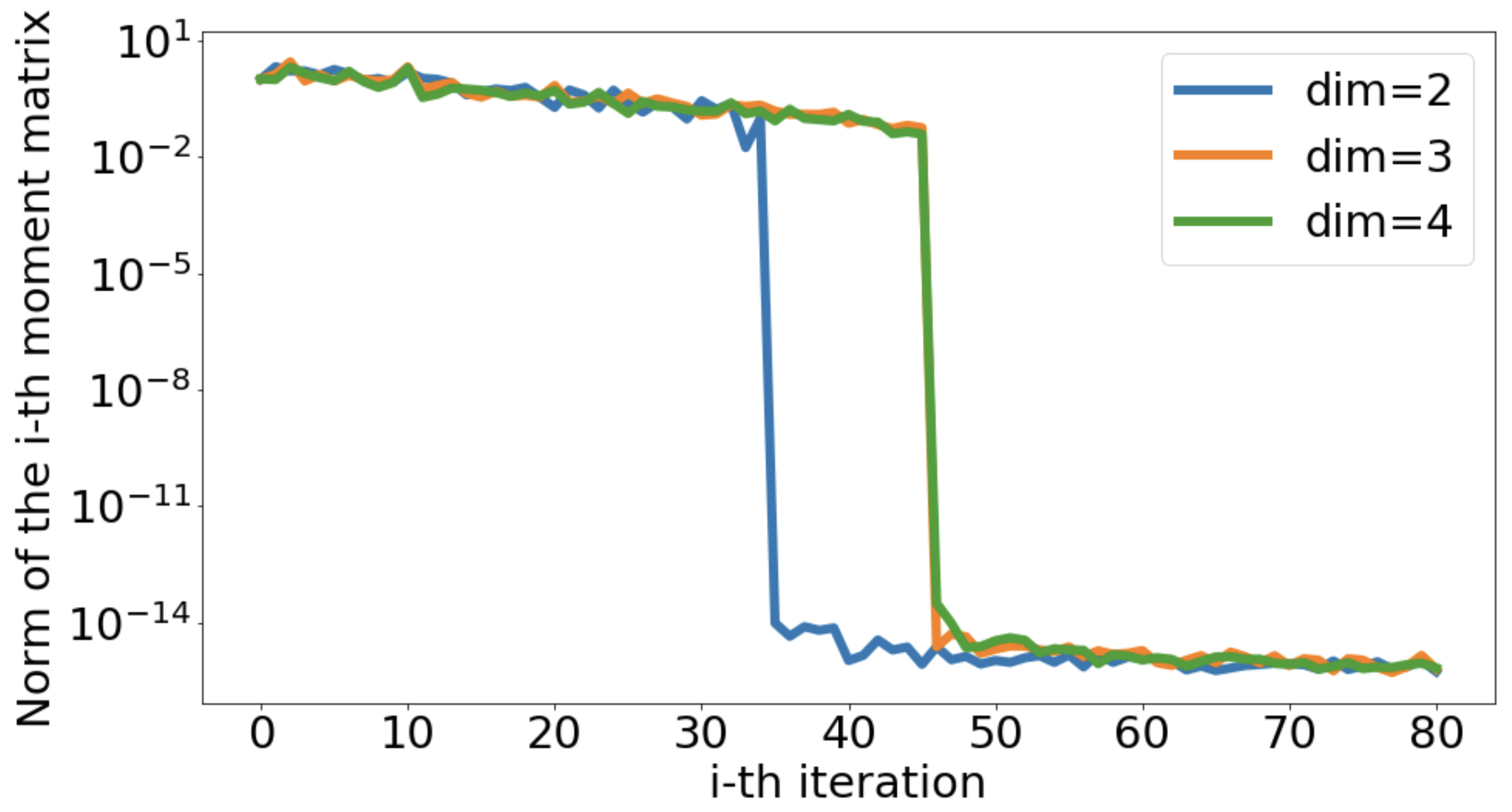}
  \caption{Norm (in log scale) of the i-th moment matrix during the Gram Schmidt process in the 3-2-2 scenario at the first level of the hierarchy and for $d\in\{2,3,4\}$. We can see that for all considered cases we observe a clear drop of the norm to zero (up to numerical artifacts) indicating when to stop the Gram Schmidt Process.}
  \label{fig:build_basis_gs}
\end{figure}

An efficient way to build a basis using the Gram Schmidt process is to generate a random moment matrix and using the standard Gram Schmidt process on this matrix with the previously generated moment matrices. By checking the norm of the matrix after removing the `projections' from the previous moment matrices at each iteration, we can decide to stop the process when only the null matrix is left. Due to numerical precision the resulting matrix will be non zero but small enough to be detected. In Fig.~\ref{fig:build_basis_gs} we show the norm of the resulted moment matrices after each iteration in the 3-2-2 scenario. We clearly see a a very large drop of the norms through the iterations. The norms drop suddenly to the order around $10^{-14}$, which is numerical noise from the computation. While we cannot guarantee the number of LI moment matrices, this is a convincing empirical evidence. Hence the number of LI moment matrices is the number of iterations just before such drop. We can clearly see that such drop does not appear all the time at the same number of iterations depending on the dimension considered. The drops for $d=2$ and $d=\{3,4\}$ are at different number of iterations, which indicates different numbers of LI moment matrices for these two cases. This important detail is discussed in Sec.~\ref{sec:ClassBasCar}.

\subsection{Classification via Basis Cardinal}\label{sec:ClassBasCar}

A basis for the linear span of the set of the moment matrices $\mathcal{M}_d^k$ can be constructed following the randomized method described in Sec.\ref{sec:bbmm}. The number of elements in the basis depends on the scenario (the MLO), the level of the hierarchy $k$ and the dimension of the Hilbert space $d$.

\begin{figure}[ht]
  \centering
  \includegraphics[width=8cm]{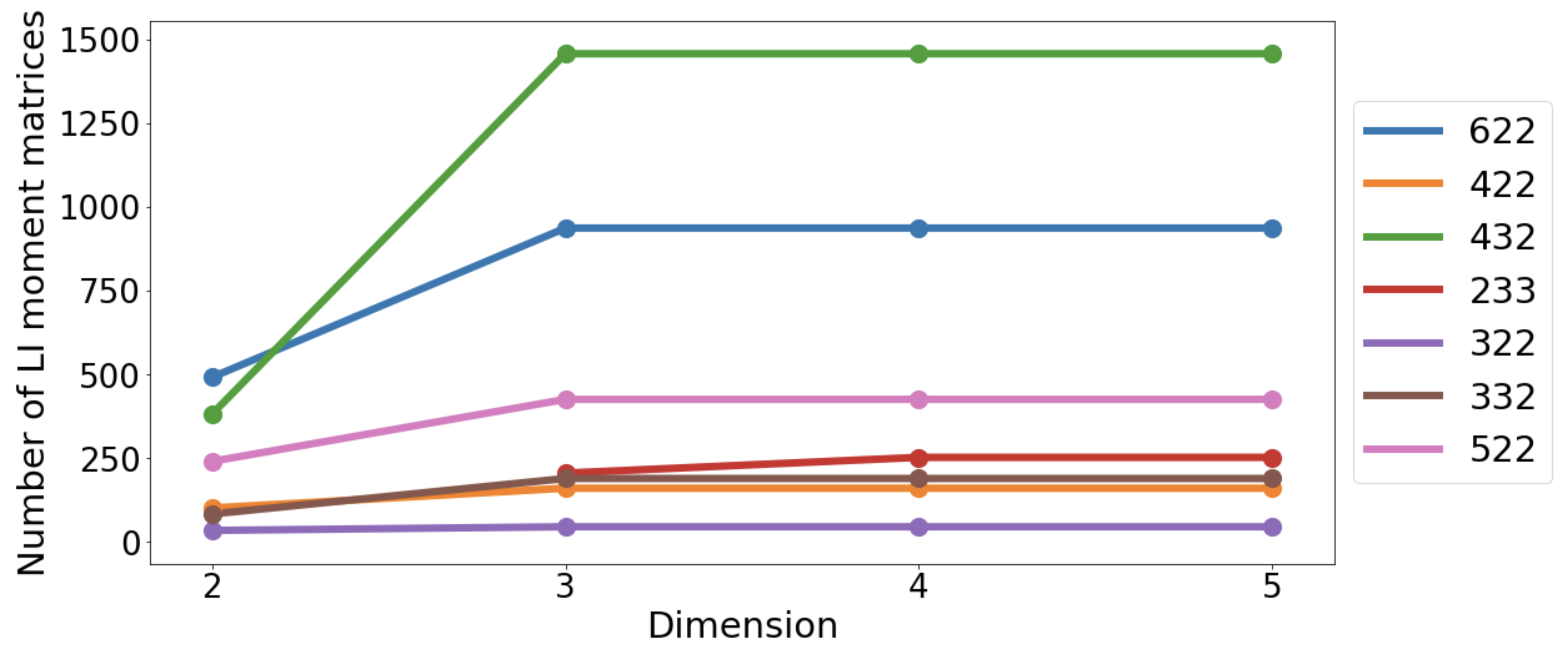}
  \caption{Maximal number of linearly independent (LI) moment matrices in the sets $\mathcal{M}_{d\in\{2,3,4,5\}}^{k=1}$. The different colors indicate the scenario described in the legend by their MLO.}
  \label{fig:Dim_CLass_MLO}
\end{figure}

The number of elements of the basis is represented in Fig.~\ref{fig:Dim_CLass_MLO} as a function of the dimension in different scenarios. In the tested scenarios with MLO $m$-$l$-2 with $m\in\{2,3,4,5,6\}$ for $l=2$ and $m\in\{2,3,4,5\}$ for $l=3$ the number of elements in the basis is the same for $d=\{3,4,5\}$ and smaller for $d=2$. This already tell us the following important information, in these considered scenarios we have the strict inclusion $\mathcal{Q}_{d=2}^{k=1} \subset \mathcal{Q}_{d>2}^{k=1}$.

\begin{figure}[ht]
  \centering
  \includegraphics[width=8cm]{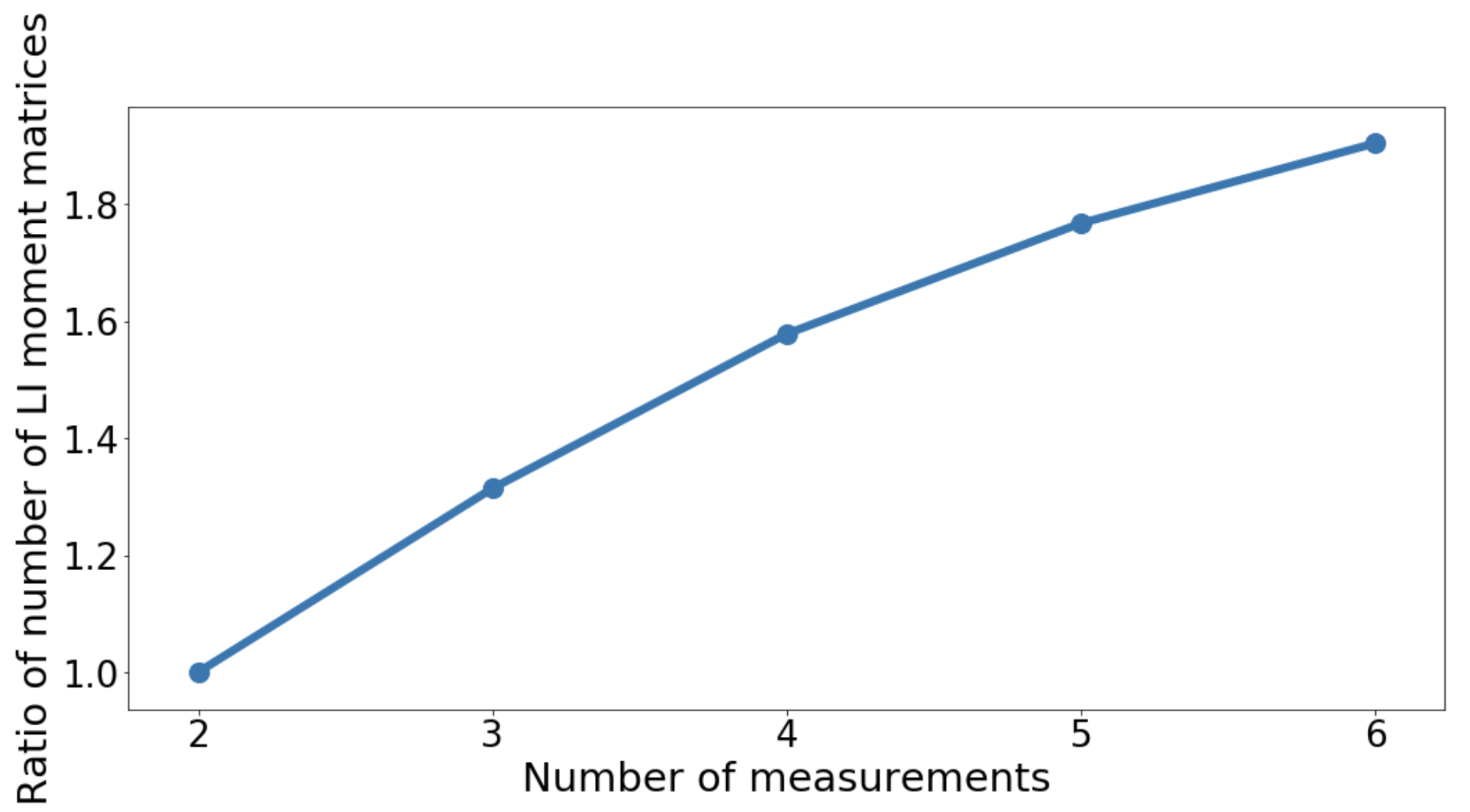}
  \caption{Ratio between the number of elements of the basis for $d=\{3,4,5\}$ (they are all equal in this case) and $d=2$ as a function of the number of measurements in the $m$-2-2 scenario.}
  \label{fig:RatioM22}
\end{figure}

To quantify this gap, the ratio between the number of elements of the basis for $d=\{3,4,5\}$ and $d=2$ is represented in Fig.~\ref{fig:RatioM22} as a function of the number of measurements in the $m$-2-2 scenario.

Another curve in Fig.~\ref{fig:Dim_CLass_MLO} corresponds to the scenario 2-3-3 for $d=\{3,4,5\}$. In this scenario we see that the number of linearly independent moment matrices for $d=3$ is smaller than the one for $d=\{4,5\}$. Hence in the 2-3-3 scenario, we have the strict inclusion $\mathcal{Q}_{d=3}^{k=1} \subset \mathcal{Q}_{d>3}^{k=1}$.

\subsection{Proposal}\label{sec:prop}

One could try to derive the basis for higher level of the hierarchy until its possible convergence. However, it is computationally expensive and potentially subject to error due to numerical precision. Another option is to use the fact that $\mathcal{Q}_{d} \subseteq \mathcal{Q}_{d}^k$. In this case, we could test whether a given behavior $\mathbf{P}$ is not in $\mathcal{Q}_{d}^k$. If it is the case then we know that $\mathbf{P} \notin \mathcal{Q}_{d}$. But if one find that $\mathbf{P}\in\mathcal{Q}_{d}^k$ this does not imply $\mathbf{P} \notin \mathcal{Q}_{d}$. In this method one can certify a minimum dimension of quantum system by its behavior without the need to characterize moment matrices sets at a computationally challenging level of the hierarchy. This correspond to take an outer approximation of the set $\mathcal{Q}$, which has shown good results using the NPA hierarchy \cite{NPA:njp08, PNA:siam10} in the context of Bell scenario \cite{PhysRevX.4.011011,PhysRevA.92.062120} but have not been used for dimension certification yet.

In addition of simply testing if a given behavior  $\mathbf{P}$ belongs to $\mathcal{Q}_{d}^k$, we are also going to quantify ``how much'' the behavior is outside the set $\mathcal{Q}_{d}^k$. This will be done by means of robustness, which is analogous to the robustness of entanglement \cite{VT:PRA98} and have also been used to quantify quantum EPR-steering \cite{CS:RPP16}, measurement incompatibility \cite{BQGMCC:PRA17,DFK:NJP19}, indefinite causality \cite{ABCFGC:NJP15}, coherence \cite{Oszmaniec2019operational,Baek_2020},and other quantities in quantum information theory.
Given a behavior $\mathbf{P}$, the general robustness visibility is

\begin{subequations}
  \label{eq:Rob}
\begin{alignat}{2}
&\text{given } &\qquad& \mathbf{P}:=\{P(\mathbf{r}|\mathbf{s})\}_{\mathbf{r},\mathbf{s}}\notag\\
\nu:= &\underset{\eta,\mathbf{P}_{d,k}}{\text{maximize}}         &\qquad& \eta \label{eq:nu}\\
&\text{subject to} &      & \eta  \mathbf{P} + (1-\eta) \mathbf{P}_{d,k} \in \mathcal{Q}_d^k ,\label{eq:Rob1}\\
&            &      & \mathbf{P}_{d,k} \in  \mathcal{Q}_d^k,\label{eq:Rob2}
\end{alignat}
\end{subequations}

we can then see that when $\nu < 1$ (represented in Fig.~\ref{fig:behaviors_hierachy}), we have $\mathbf{P} \notin \mathcal{Q}_{d}^k$. Also, if the quantum system describing the experiment is given by $\rho$, we can ensure that there exists a quantum state $\sigma$ such that for every visibility $\eta>\nu$, the noisy state $\eta \rho + (1-\eta) \sigma $ can generate behaviors which are not inside $\mathcal{Q}_{d}^k$.
Interestingly, when solving the robustness optimization problem presented in Eq.\,\eqref{eq:Rob}, one also obtains a dimension witness to certify the dimension of the given behavior and this topic will be discussed in Sec.\,\ref{sec:bi}.
Further details and a reformulation of this problem simpler for computer codes can be found at the Appendix~\ref{app:genrob}.

\section{Main Results}\label{sec:mr}
\subsection{Certifying Dimension in the $m$-$l$-2 Scenario}
\subsubsection{3-2-2 Scenario}\label{sec:322}

In order to make a step further in the understandings of the geometry of the sets, we focus here on the specific 3-2-2 scenario, namely the Leggett-Garg scenario \cite{LG:prl85} which is the simplest known sequential measurement scenario to observe quantum features. In particular, with an eye toward certifying dimensions, a legitimate question is whether in the 3-2-2 scenario there exists a behavior given by a qutrit system that cannot be reproduced by any qubit system. In other words, is there a behavior $\mathbf{P}\in \mathcal{Q}_{3}$ such that $\mathbf{P}\notin \mathcal{Q}_{2}$. We show that the answer is yes and using our method, this question can be addressed without the need to fully characterize the set of qubit's behavior, $\mathcal{Q}_{2}$, as the first level of the SDP hierarchy \cite{NV:prl15,NFAV:pra15}, $\mathcal{Q}_{2}^1$, turns out to be sufficient.

To go even further, we estimate the probability $P(\mathbf{P}\notin \mathcal{Q}_{2}^{k} | \mathbf{P}\in \mathcal{Q}_{3})$, the probability for a randomly chosen qutrit's behavior, $\mathbf{P}\in \mathcal{Q}_{3}$, to be outside the set $\mathcal{Q}_{2}^{k}$ for a given $k$ characterizing the level of the hierarchy. We evaluate the probability $P(\mathbf{P}\notin \mathcal{Q}_{2}^{k} | \mathbf{P}\in \mathcal{Q}_{d})$ for $d\in\{3, 4, 5\}$ and $k\in\{1,2\}$. In order to evaluate this probability, we first build the basis for the linear span of $\mathcal{M}_{2}^{k\in\{1,2\}}$ corresponding to the sets $\mathcal{Q}_{2}^{k\in\{1,2\}}$ with the method described in Sec.~\ref{sec:bbmm}. Then, with the same method, we also sample the sets $\mathcal{Q}_{d\in\{3,4,5\}}$ and compute the visibilities (defined in Eq.\ref{eq:Rob}) for each sampled data point. For that purpose we used CVXPY \cite{diamond2016cvxpy,agrawal2018rewriting} with the solver MOSEK \cite{mosek}. We used about 10000 points to evaluate each probability. The code is available in \cite{github}.

The probability is estimated in the following way:
\begin{align}\label{eq:probvis}
P(\mathbf{P}\notin \mathcal{Q}_{2}^{k} | \mathbf{P}\in \mathcal{Q}_{3}) \approx \frac{N(\nu < 1)}{N_{tot}},
\end{align}
where $N(\nu < 1)$ represents the number of data points with a visibility $\nu < 1$ and $N_{tot}$ is the total number of data points.

Finally all the computed probabilities are represented in the Tab.~\ref{tab:prob322}. As these probabilities are non zero, there are possibilities to find qutrit's behaviors in the 3-2-2 scenario that cannot be reproduce by any qubit's behavior. To obtain this information it is not necessary to characterize the set of qubit's behavior $\mathcal{Q}_{2}$ directly. This is because the probability $P(\mathbf{P}\notin \mathcal{Q}_{2}^{1} | \mathbf{P}\in \mathcal{Q}_{3}) \approx 0.37$ and $\mathcal{Q}_{2} \subseteq \dots \subseteq \mathcal{Q}_{2}^{2} \subseteq \mathcal{Q}_{2}^{1}$.

\begin{table}[ht]
  \centering
  \begin{tabular}{|l||c|c|c|}
    \hline
    $d$ & $k=1$ & $k=2$ \\
    \hline
		\hline
    3 & 0.365198 & 0.392267  \\
    \hline
    4 & 0.268526 & 0.317939  \\
    \hline
    5 & 0.215904 & 0.247898  \\
    \hline
  \end{tabular}
  \caption{Probability, $P(\mathbf{P}\notin \mathcal{Q}_{2}^{k} | \mathbf{P}\in \mathcal{Q}_{d})$, for a behavior in $\mathcal{Q}_{d\in\{3,4,5\}}$ to be outside $\mathcal{Q}_{2}^{k\in\{1,2\}}$.}
  \label{tab:prob322}
\end{table}

Moreover, we have $P(\mathbf{P}\notin \mathcal{Q}_{2}^{1} | \mathbf{P}\in \mathcal{Q}_{3}) < P(\mathbf{P}\notin \mathcal{Q}_{2}^{2} | \mathbf{P}\in \mathcal{Q}_{3}) \approx 0.39$. Hence, a behavior $\mathbf{P}\in \mathcal{Q}_{3}$ has more chance to be outside the second level of the hierarchy than the first level. This corresponds well to $\mathcal{Q}_{2}^{2} \subseteq \mathcal{Q}_{2}^{1}$. For all the dimensions $d \in \{3,4,5\}$, we have $P(\mathbf{P}\notin \mathcal{Q}_{2}^{1} | \mathbf{P}\in \mathcal{Q}_{d}) < P(\mathbf{P}\notin \mathcal{Q}_{2}^{2} | \mathbf{P}\in \mathcal{Q}_{d})$ as shown in Tab.~\ref{tab:prob322}.

A result that could seem counter-intuitive is when we compare the probabilities for different dimension for the same level of the hierarchy. We find that $P(\mathbf{P}\notin \mathcal{Q}_{2}^{k} | \mathbf{P}\in \mathcal{Q}_{d}) < P(\mathbf{P}\notin \mathcal{Q}_{2}^{k} | \mathbf{P}\in \mathcal{Q}_{d'})$ for $d < d'$ with $d,d' \in \{3,4,5\}$. This could sound counter-intuitive as we would expect that higher dimension could at least perform as good as the lower dimensions.

The source of these differences could be related to the sampling method. Following the discussion in Append.~\ref{app:grmm}, larger dimension means larger number of options for binning for the projectors and the use of the Haar measure to sample unitary matrices does not necessarily guarantee uniformity at the behaviors level as well. Moreover, this is related to the so-called Bertrand's Paradox \cite{Bertrand:06}, which shows that probabilities may not be well defined as they rely on the method used to produce random variables. In the Bertrand's Paradox, the method used to sample through a circle affects the probabilities. We make the analogy here, where sampling behaviors from different Hilbert spaces of different dimensions impacts the resulted probabilities. For that reasons the values presented in Tab.~\ref{tab:prob322} are not absolute and the different dimensions are not comparable. However, for a specific dimension, as the sampling is the same, the order relationship is not affected: $P(\mathbf{P}\notin \mathcal{Q}_{2}^{1} | \mathbf{P}\in \mathcal{Q}_{3}) < P(\mathbf{P}\notin \mathcal{Q}_{2}^{2} | \mathbf{P}\in \mathcal{Q}_{3})$.

The results showed in Tab.~\ref{tab:prob322} has two main consequences on our understandings of certifying dimensions of quantum systems via their behaviors. First, in the 3-2-2 scenario, the Leggett-Garg inequality \cite{LG:prl85} is already maximally violated by a qubit's behavior. However, our results implies that the 3-2-2 scenario is sufficient and inequalities based certification could be built (see Sec.~\ref{sec:bi}). Secondly, the only known way to certify qubits from the above dimensions is through the Peres-Mermin square \cite{GBCKL:pra14}, which corresponds to the 9-3-2 scenario. Our results provide a reduction of the previously known results by six measurements and shorten the length of the sequence of measurements by one which is much more favorable to experimental perspectives.

\subsubsection{Advantage to certify dimension in the m-l-2 Scenario}\label{sec:ml2}

In the previous section, in Sec.~\ref{sec:322}, we show that the scenario 3-2-2 can be used to certify some qutrit (and above dimension) behaviors that are different from those of qubits by only using the first level of the hierarchy. Moreover, as shown in Tab.~\ref{tab:prob322} using to the second level increases the chance of success. In what follow we show what are the advantage to either increase the number of measurements ($m$-2-2 scenarios) or to increase the maximum length of the sequence of measurements (3-$l$-2 scenarios).

Following the method explained in Sec.~\ref{sec:322} we evaluate the probability for a random qutrit behavior, $\mathbf{P}\in\mathcal{Q}_{3}$, not to be explained by the approximation of any qubit behavior at the first level of the hierarchy. In other words, we want to evaluate $P(\mathbf{P}\notin \mathcal{Q}_{2}^{1} | \mathbf{P}\in \mathcal{Q}_{3})$ in the scenario $m$-2-2 for $m\in \{3,\dots,8\}$.

\begin{figure}[ht]
  \centering
  \includegraphics[width=8cm]{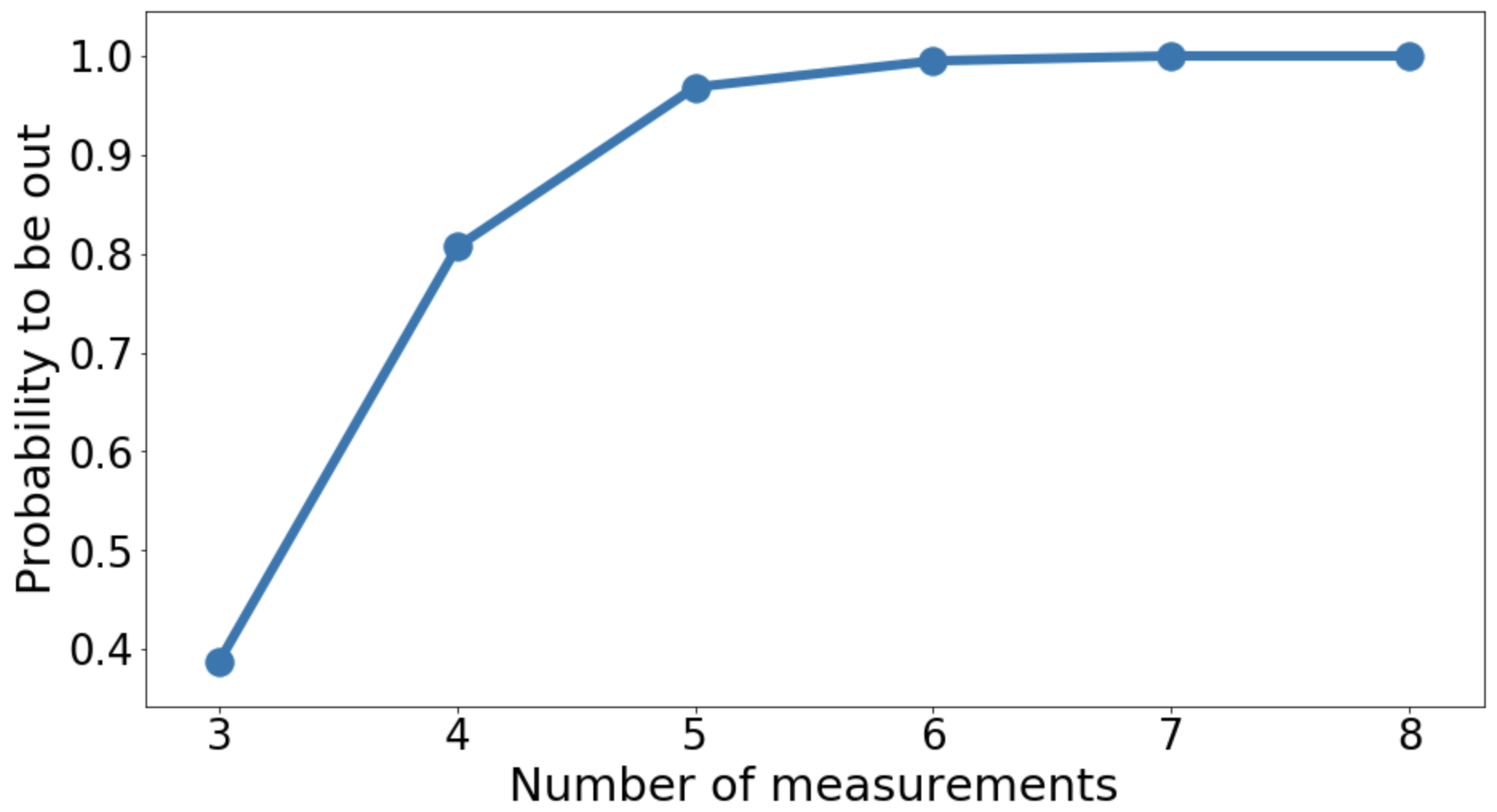}
  \caption{Probability of a random qutrit behavior not to be reproduced by a behavior in the first level of the hierarchy of qubit behaviors in the $m$-2-2 scenario.}
  \label{fig:prob_out_m22}
\end{figure}

In Fig~\ref{fig:prob_out_m22}, we show that such probability increases in a log-like manner until it reaches the value of 1 in our numerical analysis. While we observe the value 1 in our numerical analysis it is most likely that this log-like curve actually converges to 1 instead. Regardless of how close to the value 1 it is, it seems that using more than 5 or 6 measurements does not bring any further significant improvement when trying to certify dimension in this scenario. However, compared to the 3-2-2 scenario, with a probability approximately $0.37$, by using one additional measurement (4-2-2 scenario), this probability increases to approximately $0.81$, which is about the double and by using two additional measurements (5-2-2 scenario), this probability increases to approximately $0.96$, which is about the $~2.6$ times larger than in the 3-2-2 scenario.

\begin{figure}[ht]
  \centering
  \includegraphics[width=8cm]{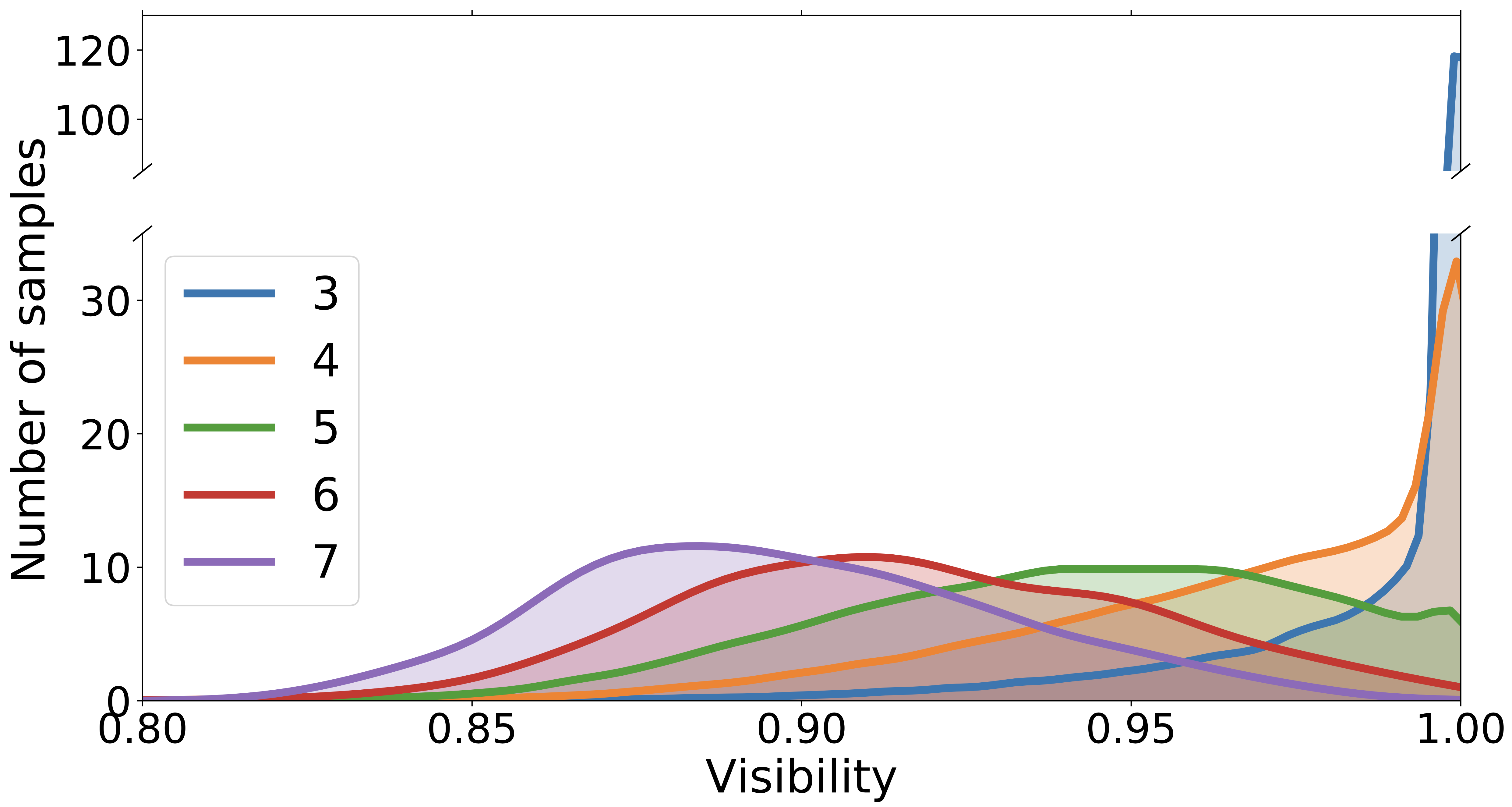}
  \caption{Critical visibility to ensure that a random qutrit behavior to be outside of the first level of the hierarchy of qubit behavior in the $m$-2-2 scenario.}
  \label{fig:vis_dist_m22}
\end{figure}

As explained in Sec.~\ref{sec:322} and in particular in Eq.\ref{eq:probvis}, the probabilities are evaluated by using the visibility. Looking at the distribution of the visibility provides also a very good insight as the visibility has a good geometrical interpretation and gives a more fine grain level of information. In Fig~\ref{fig:vis_dist_m22}, we show the distribution of the visibility of random qutrit behaviors, $\mathbf{P}\in\mathcal{Q}_{3}$ compared to the set $\mathcal{Q}_{2}^{1}$ in the scenario $m$-2-2 for $m\in \{3,\dots,8\}$. Interestingly all the distributions are different in way that when the number of measurement increases in the scenario $m$-2-2, the distribution's mean value becomes smaller. From a geometric perspective, one can say that the behaviors are further from to qubit behaviors when the number of measurement increases.

\begin{figure}[ht]
  \centering
  \includegraphics[width=8cm]{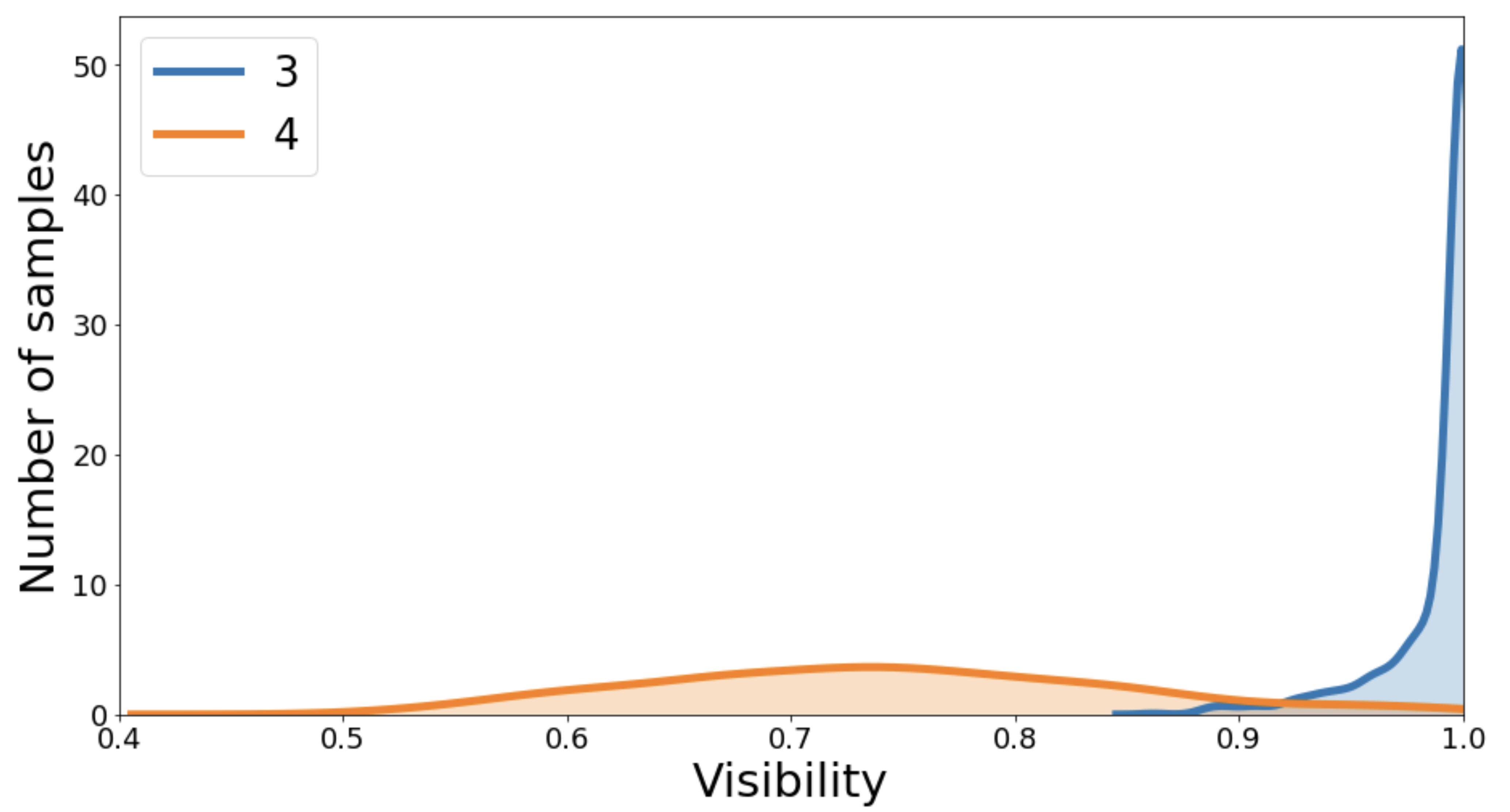}
  \caption{Critical visibility to ensure that a random qutrit behaviors is outside of the second level of the hierarchy of qubit behavior in the 3-l-2 scenario.}
  \label{fig:vist_dist_3k2}
\end{figure}

The other parameter that is possible to change experimentally is the length of the sequence of measurements used in the experimental setup. It is important to clarify that this is different compared to changing the level of the hierarchy. In the case where we increase the length experimentally, we need to collect data for the right length of sequence, while by increasing the level of the hierarchy these data would not be given. In Fig~\ref{fig:vist_dist_3k2} we show the distribution of the visibility of random qutrit behaviors, $\mathbf{P}\in\mathcal{Q}_{3}$, compared to the set of qubit behavior at the first level of the hierarchy in the scenario 3-$l$-2 for $l\in \{3,4\}$. Similarly to the number of measurements, increasing the length of the sequence of measurements makes the behavior further from to the set of qubit behaviors.

\begin{figure}[ht]
  \centering
  \includegraphics[width=8cm]{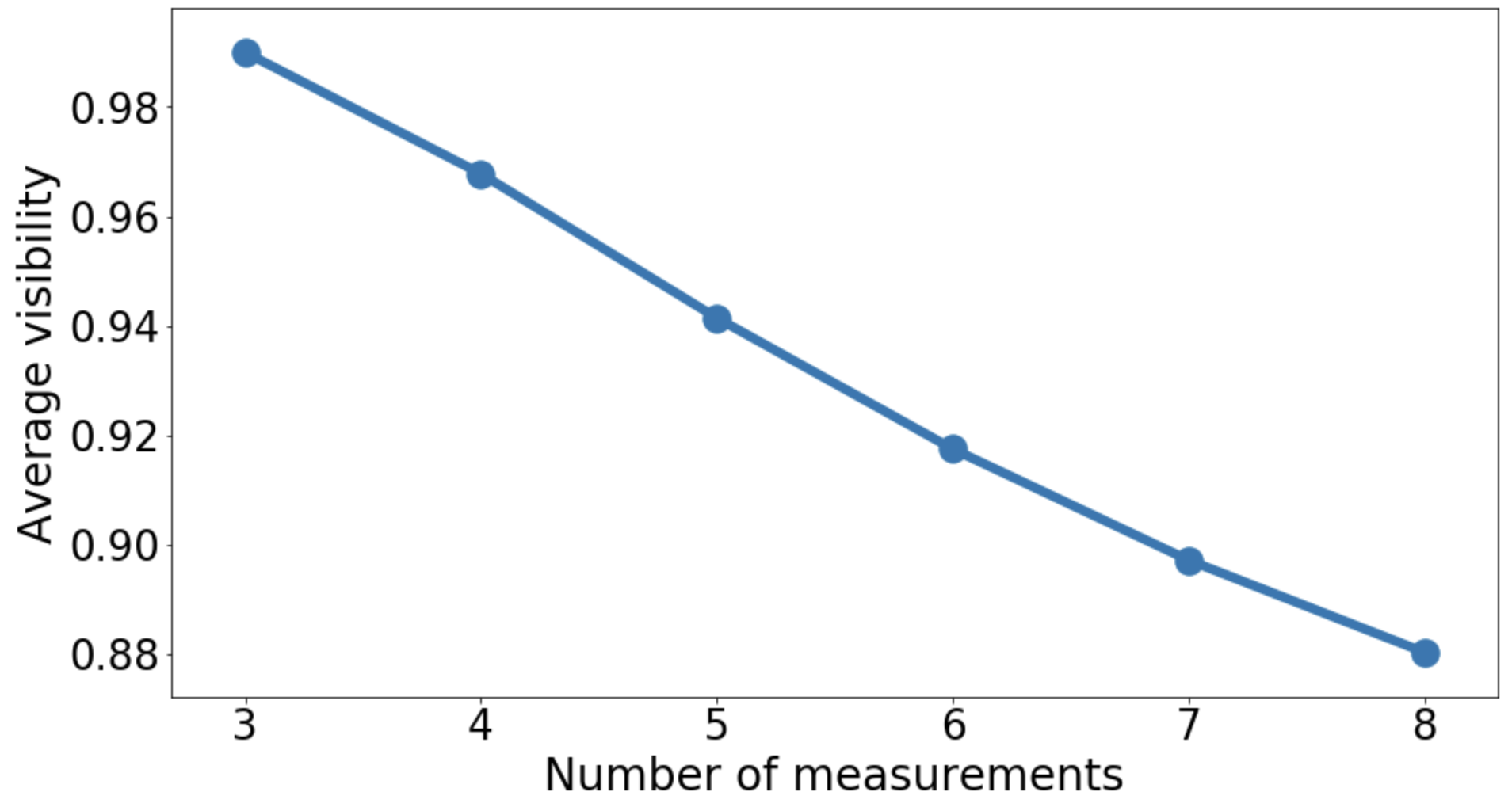}
  \caption{Average visibility of random qutrit behaviors compared to the first level of the hierarchy of qubit behavior in the $m$-2-2 scenario for $m\in\{3,\dots,8\}$.}
  \label{fig:avg_vis_m22}
\end{figure}

\begin{figure}[ht]
  \centering
  \includegraphics[width=8cm]{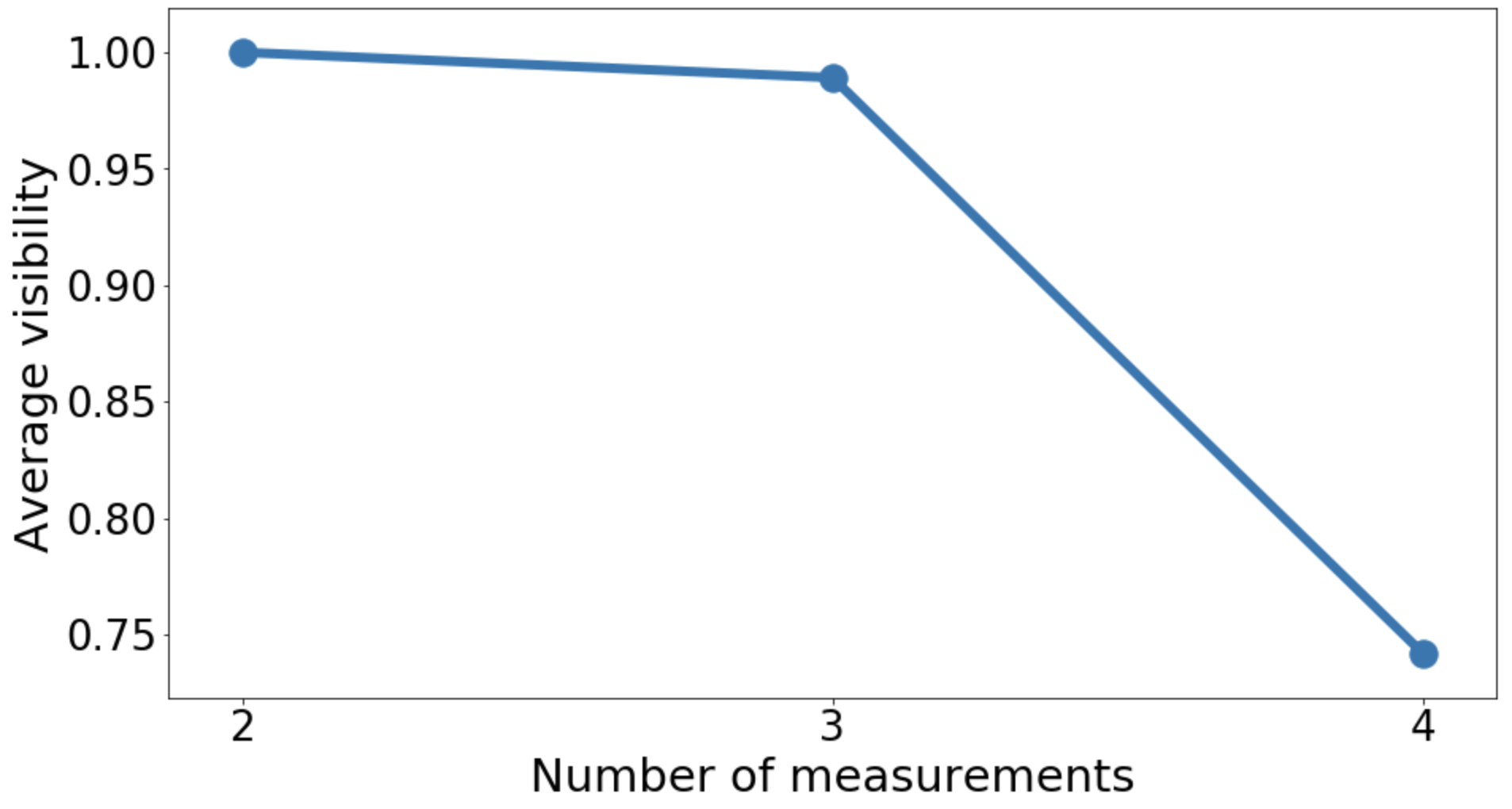}
  \caption{Average visibility of random qutrit behaviors compared to the first level of the hierarchy of qubit behavior in the 3-$l$-2 scenario for $l\in\{2,3,4\}$.}
  \label{fig:avg_vis_3k2}
\end{figure}

Our analysis shows that there are advantages to either increase the number of measurements ($m$-2-2 scenarios) or to increase the maximum length of the sequence of measurement (3-$l$-2 scenarios) to increase the chance to certify qutrit behaviors from qubit behaviors. These two parameters require a change in the experimental setup when collecting data while when by increasing the level of the hierarchy it is only about the data processing on a classical computer. While we show that increasing the number of measurements or increasing the maximum length of the sequence of measurement, there are also noticeable differences. Indeed, in the scenario 8-2-2 the average visibility is $0.88$ (see Fig.~\ref{fig:avg_vis_m22}) while in the 3-4-2 scenario the average visibility is about $0.75$. (see Fig.~\ref{fig:avg_vis_3k2}) Hence, increasing the length of the sequence of measurements seems more effective. However, this could be more challenging from an experimental point of view.

Our analysis on the distribution of the visibility also shows that increasing one of these parameters could make an experiment more robust to noise. Because of experimental precision it might sometimes not be clear whether a qutrit behavior could be obtained with a qubit system. A solution to this issue is to prefer a scenario where behaviors are not likely to be close (in a geometrical way) to qubit behaviors.
Another important result is when increasing these parameters, even with random measurements, it becomes possible to certify almost all genuine qutrit behaviors from any qubit behavior.

\subsection{Dimension Witnesses from the Robustness' Dual}
\subsubsection{Building Dimension Witness for qubits}\label{sec:bi}

Witnesses are simple to use and commonly used to test quantum properties in an experimental setup. From a geometrical perspective, they correspond to hyperplanes in the probability space. In Sec.~\ref{sec:inf} we present what we call the \emph{feasibility problem} to test whether a given behavior admits a quantum realization with a specific dimension $d$. A partial answer to this decision problem could be given by using the hyperplane in an inequality form when one ensures that all the behaviors from quantum realizations with a specific dimension $d$ do not violate the inequality as a consequence of the hyperplane separation theorem. Hence, when the inequality is violated we can infer that the behavior does not admit a quantum realization of dimension $d$.

When a behavior $\mathbf{P}'$ does not belong to the set $\mathcal{Q}^k_{d}$, one can always find a set of real coefficients $\{ \gamma_{\mathbf{r}|\mathbf{s}}\}_\mathbf{rs}$ and a real number $x$ such that
\begin{equation}\label{eq:ineq}
	\sum_{\mathbf{r},\mathbf{s}} \gamma_{\mathbf{r}|\mathbf{s}} P_{d,k}(\mathbf{r}|\mathbf{s})\geq -x, \forall \mathbf{P}_{d,k}\in \mathcal{Q}_{d}^{k},
\end{equation}
but
\begin{equation}
	\sum_{\mathbf{r},\mathbf{s}} \gamma_{\mathbf{r}|\mathbf{s}} P'(\mathbf{r}|\mathbf{s})< -x .
\end{equation}

Moreover such an inequality can be explicitly obtained by means of the dual formulation of the robustness optimization problem presented in Eq.~\eqref{eq:Rob}. We present the details on the dual formulation and how to construct such inequalities in the Appendix~\ref{app:genrob}. Also, to demonstrate this technique, we provide an explicit example (available at the github repository \cite{github}) in the scenario 3-2-2, where we obtained a randomly generated behavior from a qutrit and solve the generalized robustness problem. The inequality we obtain can be used in any experiment in the scenario 3-2-2 to test if the experimentally obtained behavior does not admit a quantum realization with a qubit.

\subsection{Higher Dimension Witness}

As it has been shown in Sec.~\ref{sec:ClassBasCar} and in Fig.~\ref{fig:Dim_CLass_MLO}, in the 2-3-3 scenario, we have $\mathcal{Q}_{d=3}^{k=1} \subseteq \mathcal{Q}_{d>3}^{k=1}$. This suggests that there might be ququart behaviors that cannot be obtained via qutrit behaviors as we have seen in the case of dimension higher than two in the m-l-2 scenario presented in Sec.\ref{sec:ml2}. Following the same method used in Sec.\ref{sec:bi}, we provide an explicit example of both a dimension witness and quantum realization (ququart quantum state and quantum measurements) (available at the github repository \cite{github}) whose behavior cannot be reproduced by qutrits. This is done by generating a random ququart behavior in the 2-3-3 scenario and computing its visibility regarding the set $\mathcal{Q}_{d=3}^{1}$. Once such behavior is found we can find the dimension witness by using the dual problem of the generalized robustness.

\section{Conclusion}

In this work, we have refined the known methods that analyze correlations arising from quantum systems subjected to sequential projective measurements to certify that the system in question necessarily has a quantum dimension greater than some dimension $d$. We have shown that, in the context of dimension certification, it is not necessary to go to high level in the NV hierarchy and in all the examples we have treated the first level of the hierarchy was already sufficient to certify dimensions.

We have shown that scenarios can be classified by the number of elements in the moment matrices basis at the first level of the hierarchy. For all the scenarios where different number of basis elements have been recorded, we have shown that dimension can be certified in these scenarios. While this seems convincing to say that this property is a necessary condition, it is still an open problem. Moreover, we demonstrated that the randomized method to build the basis of moment matrices that was previously considered as a non accurate method due to potential numerical precision can actually be accurate with empirical evidence if one keeps track of the moment matrices norms in the Gram Schmidt process. We showed how dimension witnesses can be obtained from our method and we have provided concrete numerical examples.

With this method we found that in the 3-2-2 scenario, while the known Leggett-Garg inequality is already maximally violated by a qubit's behavior our results implies that the 3-2-2 scenario is sufficient. The only previously known way to certify qubit from the above dimensions was through the Peres-Mermin square, which corresponds to the 9-3-2 scenario in our notation. Our results provide a drastic simplification of the previously known results by six measurements and shorten the length of the sequence of measurements by one which is much more favorable to experimental perspectives.
We provided a deep analysis and a characterization of scenarios in which dimension certification of dimension greater than two is possible. In particular, we use two metrics: the probability for a random quantum states with random measurements that does not admit any qubit quantum realization and the distribution of the visibility derived from the generalized robustness. We observed that the probability to find a behaviors with no qubit quantum realization increases when the experiment is more complex (more measurements or longer sequence of measurements). This is also possible by increasing the level of the hierarchy. When we increase the number of measurements or the length of the sequence we see that the distribution of the visibility shifts to lower values, this can be interpreted as a more robust certification. In summary, in the selection of scenarios, when the experiment is more complex there are more chances to make a successful certification which is also getting more robust.

In addition, our methods led us to certify, for the first time in this projective measurements scenario, quantum systems with dimensions strictly greater than three.

Even though there are important results for scenarios considering general measurements \cite{Hoffmann_2018,Spee_2020, mao2020structure}, a systematic approach like the one presented here for the projective case is still missing for the general measurements case. Unfortunately, with the method used here, when we lower bound a dimension from its behavior, it only correspond to behaviors arising from projective measurements. Projective measurements seem to be a crucial hypothesis for moment matrices methods as Eq.~\ref{eq:moment_condition} does not hold without assuming projective measurements. We notice that the projective measurement assumption is also required in the standard NPA hierarchy \cite{NPA:njp08, PNA:siam10}, the finite dimension NV hierarchy \cite{NV:prl15,NFAV:pra15}, and the moment matrix approach used by Budroni-Moroder-Kleinmann-Gühne \cite{BMKG:prl13}. It is an open problem how this could be resolved in general measurement scenarios. One possible solution to enforce the use of the moment matrices based method, would be to use it together with self-testing methods \cite{_upi__2020} which has already been used for this purpose in the prepare-and-measure and contextuality scenario \cite{PhysRevLett.122.250403}.

We have implemented the methods described in this paper using the language Python and all code is publicly available on the online repository \cite{github} and free to use under the Apache License 2.0.

\section*{Acknowledgements}

AS and JK have been supported by a KIAS individual grant (CG070301 and CG014604) at Korea Institute for Advanced Study. DM acknowledges support of the ANR through the ANR-17-CE24-0035 VanQuTe. AS would like to thank S. Akibue. MTQ would like to thank M. Ara\'ujo. MTQ acknowledges support of the Austrian Science Fund (FWF) through the SFB project "BeyondC", a grant from the Foundational Questions Institute (FQXi) Fund and a grant from the John Templeton Foundation (Project No. 61466) as part of the The Quantum Information Structure of Spacetime (QISS) Project (qiss.fr). The opinions expressed in this publication are those of the authors and do not necessarily reflect the views of the John Templeton Foundation. The authors are grateful to the Center for Advanced Computation at Korea Institute for Advanced Study for help with computing resources.


\bibliographystyle{apsrev4-2.bst}
\bibliography{biblio}

\onecolumn\newpage
\appendix
\section*{{Appendix}}

\section{Optimization of quantum realization for finite dimension and under polynomial constraints}\label{app:polyopt}

Generally, the \emph{optimization problem} can be represented by what follows

\begin{subequations}
  \label{eq:optdim}
\begin{alignat}{2}
p_{d}^* := &\underset{\mathcal{H}, \mathbf{X}, \rho}{\text{maximize}}    &\qquad& \text{Tr}[ p(\mathbf{X}) \rho ]\label{eq:polyXmax}\\
&\text{subject to} &      & \dim(\mathcal{H}) \leq d ,\label{eq:dimconst}\\
&                  &      & q_i(\mathbf{X}) \succeq 0, \forall i \in \{1,\dots,q\},\label{eq:polyconst}
\end{alignat}
\end{subequations}
where $\mathbf{X}$ is a set of observables, $p(.)$ and $q_i(.)$ are polynomial functions and $d$ and $q$ are both integers. The $X_i$ are not to be confused as the moment matrix $M$. While they are different, they are connected as the moment matrix is build from the expected values of the obsevables' projectors.

This is more general than the case we actually need to consider here as we might not impose any constraints represented by the $q_i(.)$ polynomials. However, the optimization problem described in Eq.~\ref{eq:optdim} is regularly encountered in quantum information. For instance, in contextuality, one of the simplest known inequality is the so-called KCBS inequality \cite{KCBS:prl08}. It resembles very much to a sequential measurement scenario as we consider with an MLO of 5-2-2, where we impose commutation relationships as $[X_i,X_{i+1}]=0$.

\begin{subequations}
  \label{eq:optdimKCBS}
\begin{alignat}{2}
p_{d,KCBS}^* := &\underset{\mathcal{H}, \mathbf{X}, \rho}{\text{maximize}}     &\qquad& \text{Tr}[ \sum_{i=1}^5 X_i X_{i+1} \rho ]\label{eq:KCBSpolyXmax}\\
&\text{subject to} &      & \dim(\mathcal{H}) \leq d ,\label{eq:KCBSdimconst}\\
&                  &      & [X_i,X_{i+1}]=0. \label{eq:KCBSpolyconst}
\end{alignat}
\end{subequations}

While this optimization program has a clear interpretation it is not straight forward to solve it. However, it is possible to re-express it using the moment matrix representation to solve the optimization problem via a hierarchy of semidefinite programming relaxations \cite{NV:prl15,NFAV:pra15} based on the original NPA hierarchy \cite{NPA:njp08, PNA:siam10}. The $k$-th level of the hierarchy has a SDP formulation.

\section{Guess Your Neighbor's Input Inequality}\label{app:gyni}

In Sec.~\ref{sec:fin}, we provided a specific example to show that moment matrices do not necessarily have a quantum realization. Moreover, in this specific example there exist no quantum realization. It is in the case of the so-called guess-your-neighbor’s-input inequality \cite{ABBAGP:prl10}, in particular in its sequential measurement scenario \cite{BMKG:prl13}. It is a sequential scenario with a MLO of 2-3-2 and the inequality is:
\begin{align}
  P(000|000) + P(110|011) + P(011|101) +P(101|110) \leq 1.
\end{align}
 As mentioned in Sec.~\ref{sec:fin}, by solving the SDP referred in Eq.~\ref{eq:SDPdiminf} one can find the maximum value $p^* \approx 1.0225$ when the dimension is unrestricted \cite{BMKG:prl13}.

We are interested to know what is the maximum for the behaviors $\mathbf{P} \in \mathcal{Q}_{d=2}^{k=1}$. This can be done by solving the SDP in Eq.~\ref{eq:SDPdimfin}. In this case it is:
\begin{subequations}
  \label{eq:SDP_GYNI}
\begin{alignat}{2}
p_{GYNI,d=2,k=1}^* := &\underset{M}{\text{maximize}}        &\qquad& \sum_{\mathbf{r},\mathbf{s}} \gamma_{GYNI,\mathbf{r}\vert\mathbf{s}} M_{\mathbf{r}\vert\mathbf{s}}\label{eq:Mmaxk_gyni}\\
&\text{subject to} &      & M_{\mathbf{0}\vert\mathbf{0},\mathbf{0}\vert\mathbf{0}} = 1 ,\label{eq:Mid_gyni}\\
&            &      & M \in \mathcal{M}_{d=2}^{k=1},\label{eq:Mdkgyni}\\
&            &      & M \succeq 0,\label{eq:Mpos_gyni}
\end{alignat}
\end{subequations}
where
\begin{align}\label{eq:coeff_GYNI}
\gamma_{GYNI,\mathbf{r}\vert\mathbf{s}} = \left\{
    \begin{array}{ll}
        1 & \mbox{if } \mathbf{r}\vert\mathbf{s} \in E_{GYNI} \\
        0 & \mbox{otherwise.}
    \end{array}
\right.
\end{align}
and $E_{GYNI} = \{(000|000),(110|011),(011|101),(101|110)\}$ is the set of events appearing in the inequality in Eq.\ref{eq:gyni}. The final form of the SDP is then
\begin{subequations}
  \label{eq:SDP_GYNIv2}
\begin{alignat}{2}
p_{GYNI,d=2,k=1}^* := &\underset{M}{\text{maximize}}        &\qquad& \sum_{\mathbf{r},\mathbf{s}} \gamma_{GYNI,\mathbf{r}\vert\mathbf{s}} M_{\mathbf{r}\vert\mathbf{s}}\label{eq:Mmaxk_gyniv2}\\
&\text{subject to} &      & M_{\mathbf{0}\vert\mathbf{0},\mathbf{0}\vert\mathbf{0}} = 1 ,\label{eq:Mid_gyniv2}\\
&            &      & M = \sum_{i=1}^n \alpha_i M_{d=2,i}^{k=1},\label{eq:Mdk_gyniv2}\\
&            &      & \mathbf{\alpha}\in \mathbb{R}^n,\\
&            &      & M \succeq 0,\label{eq:Mpos_gyniv2}
\end{alignat}
\end{subequations}
where $\{M_{d=2,i}^{k=1}\}_{i=1}^n$ is a basis of $\mathcal{M}_{d=2}^{k=1}$.

We obtain $p_{2,1}^* \approx 1.1588$, showing that in the scenario 2-3-2, some behaviors $\mathbf{P} \in \mathcal{Q}_{d=2}^{k=1}$ do not admit any quantum realization. The code is provided in \cite{github}.

\section{Generating Random Moment Matrices}\label{app:grmm}

In what follow we refer to the randomized method used in \cite{NFAV:pra15}. A moment matrix, $M$, can be obtained by the Eq.~\ref{eq:moment_matrix} if a quantum state $\rho$ and a set of projectors $\{\Pi_{{r_i}\vert s_i}\}$ is provided. In the randomized method, we randomly select a state form the space of states and each projective measurements in the set of projectors. They are multiple ways to do this and we have adopted the following one. The state $\rho$ is prepared as follows:
\begin{align}
  \rho = U \vert 0 \rangle \langle 0 \vert U^\dagger,
\end{align}
where $U$ is a random $d\times d$ unitary matrix obtained using the \textit{Haar measure} \cite{FM:nams07}.

For the projectors, it is more subtle as we need to satisfy completeness: $\sum_{r_i}\Pi_{r_i}\vert{s_i} = \mathbb{1}$ for every $s_i$. When the number of outcomes is equal to the dimension of the Hilbert space, in order to satisfy the normalization and orthogonality among projectors of different outcomes, all the projectors must be rank-1. However, when the dimension is larger than the number of outcomes some of the projectors must have higher rank. The assignment of the rank of the projectors can be done using a pseudo-random number generator such as the Mersenne Twister \cite{MN:acm98}. In other words, this corresponds to a binning method. To assign each result $r_i\in R_i$ to different dimension (to keep the orthogonality), we can define the different sets $B_{r_i | s_i}$ $\forall r_i \in R_i$ such that: $B_{r_i | s_i}\succeq \{0,\dots,d-1\}$, $B_{r_i | s_i}\cap B_{r_i' | s_i} = \emptyset$ if $r_i \neq r_i'$ and $\cup_{r_i \in R_i} B_{r_i | s_i} = \{0,\dots,d-1\}$.

The projector $\Pi_{r_i\vert s_i}$ is prepared as follows:
\begin{align}
  \Pi_{r_i\vert s_i} =  \sum_{j \in B_{r_i | s_i}} U_{s_i}\vert j \rangle \langle j \vert U_{s_i}^\dagger,
\end{align}
where $U_{s_i}$ is a random $d\times d$ unitary matrix obtained using the \textit{Haar measure} \cite{FM:nams07} and characteristic of the setting $s_i$. It is important to keep the same unitary $U_{s_i}$ for all the projectors with the same setting in order to keep the orthogonality. Then each projectors of sequences of measurements can be obtained as explained in Sec.~\ref{sec:sms} with the expression $\Pi_{\mathbf{r}\vert\mathbf{s}} = \Pi_{r_l\vert s_l}\dots\Pi_{r_1\vert s_1}$.

A behavior $\mathbf{P}$ can also be randomly created using the same technique as the diagonal elements of a moment matrix obtained by this technique is a behavior.

\section{Generalized Robustness}\label{app:genrob}

In Sec.~\ref{sec:prop} (in particular in Eq.~\ref{eq:Rob}) we have presented an optimization problem to quantify ``how far'' is a given behavior $\mathbf{P}:=\{P(\mathbf{r}|\mathbf{s})\}_{\mathbf{r},\mathbf{s}} $ from the set  $\mathcal{Q}_d^k$. This quantifier is analogous to the robust of entanglement\footnote{We recommend Ref.~\cite{CS:RPP16} for an introduction on robust quantifiers and SDP.} \cite{VT:PRA98} and we defined it as:
\begin{subequations}
\begin{alignat}{2}
&\text{given } &\qquad& \mathbf{P}:=\{P(\mathbf{r}|\mathbf{s})\}_{\mathbf{r},\mathbf{s}}\notag\\
\nu:= &\underset{\eta,\mathbf{P}_{d,k}}{\text{maximize}}         &\qquad& \eta\notag\\
&\text{subject to} &      & \eta  \mathbf{P} + (1-\eta) \mathbf{P}_{d,k} \in \mathcal{Q}_d^k, \notag\\
&            &      & \mathbf{P}_{d,k} \in  \mathcal{Q}_d^k,
\end{alignat}
\end{subequations}
where $d$ is the dimension and $k$ the level of the hierarchy.

By generating a basis $\{M_i\}_i$ for the linear span of $\mathcal{M}_d^k$ the above optimization problem can be phrased as an SDP:
\begin{subequations}
  \label{eq:primal}
  \begin{alignat}{2}
  &\text{given } &\qquad& \{M_i\}_i, \{P(\mathbf{r}|\mathbf{s})\}_{\mathbf{r},\mathbf{s}}\notag\\
  \nu:= &\underset{\eta,X,R,\{\alpha_i\}_i,\{\beta_i\}_i}{\text{maximize}}         &\qquad& \eta\notag\\
  &\text{subject to} &      & \eta  P(\mathbf{r}|\mathbf{s}) + R_{\mathbf{r}\vert\mathbf{s},\mathbf{r}\vert\mathbf{s}} = X_{\mathbf{r}\vert\mathbf{s},\mathbf{r}\vert\mathbf{s}}, \notag\\
  &            &      & X = \sum_i \alpha_i M_i,\\
  &            &      & R = \sum_i \beta_i M_i	\\
  &            &      & X_{0|0,0|0} = 1	\\
  &            &      & R_{0|0,0|0} = 1-\eta \\
  &            &      & X\geq0, \quad R\geq0,
  \end{alignat}
\end{subequations}
where $\alpha_i$ and $\beta_i$ are real numbers and $X$ and $R$ are matrices of the size of $M_i$.
	Following the steps of Ref.\cite{boyd}, the Lagrangian of this optimization problem can then be written as
	\begin{align}
	L &= \eta + \sum_{\mathbf{r}|\mathbf{s}}   \gamma_{\mathbf{r}|\mathbf{s}}  \left(R_{\mathbf{r}\vert\mathbf{s},\mathbf{r}\vert\mathbf{s}} - X_{\mathbf{r}\vert\mathbf{s},\mathbf{r}\vert\mathbf{s}} + \eta  P(\mathbf{r}|\mathbf{s})   \right)
		+	\tr\left(A \left[X - \sum_i \alpha_i M_i \right] \right)	 \\
		&+	\tr\left(B \left[R - \sum_i \beta_i M_i \right] \right) + \tr(\rho X) + \tr(\sigma R)
		+ r( 1-\eta - R_{\mathbf{0}\vert\mathbf{0},\mathbf{0}\vert\mathbf{0}}) + x (1 - X_{\mathbf{0}\vert\mathbf{0},\mathbf{0}\vert\mathbf{0}}) \nonumber \\ \nonumber \phantom{s} \\
		&= \eta \left( 1 - r + \sum_{\mathbf{r}|\mathbf{s}}   \gamma_{\mathbf{r}|\mathbf{s}} P(\mathbf{r}|\mathbf{s}) \right)
		+ \tr \left( X \left[ \rho - x \ketbra{0|0}{0|0} + \sum_{\mathbf{r}|\mathbf{s}}   \gamma_{\mathbf{r}|\mathbf{s}}  \ketbra{\mathbf{r}\vert\mathbf{s}}{\mathbf{r}\vert\mathbf{s}}  \right]  \right) \\ \nonumber
		&+ \tr \left( R \left[ \sigma - x \ketbra{0|0}{0|0} + \sum_{\mathbf{r}|\mathbf{s}}   \gamma_{\mathbf{r}|\mathbf{s}}  \ketbra{\mathbf{r}\vert\mathbf{s}}{\mathbf{r}\vert\mathbf{s}} \right] \right)
		- \sum_i \alpha_i \tr(M_i A) -  - \sum_i \beta_i \tr(M_i B) + r +x \nonumber
	\end{align}
	for dual variables $\gamma$, $A, B, \rho, \sigma, r, $ and $x$. The dual program of the SDP described in Eq.\,\ref{eq:primal} can then be written as
\begin{subequations}
  \label{eq:dual}
  \begin{alignat}{2}
  &\text{given } &\qquad& \{M_i\}_i, \{P(\mathbf{r}|\mathbf{s})\}_{\mathbf{r},\mathbf{s}}\notag\\
  \nu:= &\underset{x,r,\{\gamma_{\mathbf{r}|\mathbf{s}}\}_{\mathbf{r}\mathbf{s}},A,B}{\text{minimize}}         &\qquad& x + r \notag\\
  &\text{subject to} &      & r = 1 + \sum_{\mathbf{r},\mathbf{s}} \gamma_{\mathbf{r}|\mathbf{s}} P(\mathbf{r}|\mathbf{s}), \notag\\
  &            &      & \sum_{\mathbf{r},\mathbf{s}}  \ketbra{\mathbf{r}\vert\mathbf{s}}{\mathbf{r}\vert\mathbf{s}} \gamma_{\mathbf{r}\vert\mathbf{s}} \geq \phantom{-} A - x \ketbra{0|0}{0|0},\\
  &            &      & \sum_{\mathbf{r},\mathbf{s}} \ketbra{\mathbf{r}\vert\mathbf{s}}{\mathbf{r}\vert\mathbf{s}} \gamma_{\mathbf{r}\vert\mathbf{s}}  \leq -B + r\ketbra{0|0}{0|0}	\\
  &            &      & \tr(M_i A) = 0, \; \forall i,	\\
  &            &      & \tr(M_i B) = 0, \; \forall i,
  \end{alignat}
\end{subequations}
where $A$ and $B$ are are matrices of the size of $M_i$, $\gamma_{\mathbf{r}|\mathbf{s}}$ $x$ and $r$ are real numbers,  $\ketbra{\mathbf{r}\vert\mathbf{s}}{\mathbf{r}\vert\mathbf{s}}$ is a matrix of the size of $M_i$ which has value $1$ on the component $(\mathbf{r}\vert\mathbf{s}, \mathbf{r}\vert\mathbf{s})$ and zero everywhere else.
Moreover, the real coefficients $\{\gamma_{\mathbf{r}|\mathbf{s}} \}_{\mathbf{r},\mathbf{s}}$ provide an inequality that can be used to certify that the behavior. $\{P(\mathbf{r}|\mathbf{s})\}_{\mathbf{r},\mathbf{s}}$ does not admit a quantum realization of dimension $d$.

More precisely, after solving the optimization problem \ref{eq:dual}, the given behavior satisfies $\sum_{\mathbf{r},\mathbf{s}} \gamma_{\mathbf{r}|\mathbf{s}} P(\mathbf{r}|\mathbf{s})=\nu-x-1$. From the primal, we see that every behavior $\{P_\mathbf{X}(\mathbf{r}|\mathbf{s})\}_{\mathbf{r},\mathbf{s}}\in \mathcal{Q}_d^k $ respects $\nu\geq1$, hence we have the witness $\sum_{\mathbf{r},\mathbf{s}} \gamma_{\mathbf{r}|\mathbf{s}} P_\mathbf{X}(\mathbf{r}|\mathbf{s})\geq -x $.
Note that if the given behavior is not inside $\mathcal{Q}_d^k$ the witness is always violated by this behavior. The primal formulation ensures that if $\{P(\mathbf{r}|\mathbf{s})\}_{\mathbf{r},\mathbf{s}}\notin \mathcal{Q}_d^k$ we have $\nu<1$, hence $\sum_{\mathbf{r},\mathbf{s}} \gamma_{\mathbf{r}|\mathbf{s}} P(\mathbf{r}|\mathbf{s})<-x$.

\end{document}